\documentclass[aps,prd, nofootinbib,preprintnumbers]{revtex4}%
\usepackage{subcaption}
\usepackage{amsmath,amsfonts,amssymb,graphicx,graphics,color, hyperref}
\usepackage[latin9]{inputenc}
\usepackage{pifont}
\usepackage{natbib}
\usepackage{epsfig,epstopdf}
\usepackage{bm}
\usepackage{enumerate}
\usepackage{graphicx}
\usepackage{amsmath}
\usepackage{amsfonts}
\usepackage{amssymb}%
\providecommand{\U}[1]{\protect\rule{.1in}{.1in}}
\newcommand{\be}{\begin{equation}}
\newcommand{\ee}{\end{equation}}
\newcommand{\bea}{\begin{eqnarray}}
\newcommand{\eea}{\end{eqnarray}}

\begin{document}
\title{Partial decidability protocol for the Wang tiling problem from statistical
mechanics and chaotic mapping}
\author{Fabrizio Canfora$^{1,2}$}
\author{Marco Cedeño$^{2}$ }
\affiliation{$^{1}$Centro de Estudios Científicos (CECs), Avenida Arturo Prat 514,
Valdivia, Chile.}
\affiliation{$^{2}$Facultad de Ingeniería, Universidad San Sebastian, sede Valdivia,
General Lagos 1163, Valdivia 5110693, Chile.}
\email{fabrizio.canfora@uss.cl, marco.cedeno@uss.cl }

\begin{abstract}
We introduce a \textit{partial decidability protocol} for the Wang tiling
problem (which is the prototype of undecidable problems in combinatorics and
statistical physics) by constructing a suitable mapping from tilings of finite
squares of different sizes. Such mapping depends on the initial family of Wang
tiles (the "alphabet") with which one would like to tile the plane. This
allows to define effective entropy and temperature associated to the alphabet
(together with the corresponding partition function). We identify a subclass
of \textit{good alphabets} by observing that when the entropy and temperature
of a given alphabet are "well-behaved" in the thermodynamical sense then such
alphabet is a good candidate to tile the infinite two-dimensional plane. Our
proposal is tested successfully with the known available good alphabets (which
produce periodic tilings, aperiodic but self-similar tilings as well as
tilings which are neither periodic nor self-similar). Our analysis shows that
the Kendall's Tau coefficient is able to distinguish alphabets with a good
thermodynamical behavior from alphabets with bad thermodynamical behavior. The
transition from good to bad behavior is related to a transition from
non-chaotic to chaotic regime in discrete dynamical systems of logistic type.

\end{abstract}
\maketitle
\tableofcontents

\section{Introduction}

The Wang tiling problem \cite{[1]} is one of the deepest and most relevant
problem in statistical physics, mathematics and computer science. A Wang tile
is a square with one color for each side. Given a finite family ("alphabet")
of Wang tiles (from now on, a Wang tile will simply be denoted as tile), the
problem is to understand whether or not one can cover the plane using these
tiles\footnote{The rule of the game is that one can use translated copies of
these tiles in such a way that two tiles can be put together along one side if
and only if the common side of two neighbor tiles has the same color in both.
Moreover, one can only use the tiles of the initial family but one cannot
rotate the tiles.} and satisfying certain rules. Here it is the original
formulation \cite{[1]}:

\textquotedblleft Assume we are given a finite set of square plates of the
same size with edges colored, each in a different manner. Suppose further
there are infinitely many copies of each plate (plate type). We are not
permitted to rotate or reflect a plate. The question is to find an effective
procedure by which we can decide, for each given finite set of plates, whether
we can cover up the whole plane (or, equivalently, an infinite quadrant
thereof) with copies of the plates subject to the restriction that adjoining
edges must have the same color.\textquotedblright\ 

This problem hides some of the most important open questions both in
mathematics and physics. As Wang himself noticed, the existence of aperiodic
tile sets is closely related to this decision problem. Berger \cite{[2]}
(which was Wang student) established the undecidability of the problem a
result which implies that there must exist aperiodic tiles set.

A small remark is in order: the sentence "to solve the Wang tiling problem"
actually means "to find an algorithm that, when receiving the input (the
initial family of tiles, the alphabet) produces the corresponding output
(which is \textbf{yes} if the alphabet can cover up the whole plane and is
\textbf{no} if it cannot)". A problem of this type is called \textit{decision
problem} (since the answer is either yes or not). A decision problem is called
undecidable if no algorithm can solve it \textit{in all instances}. In the
present case, the undecidability of the Wang tiling means that \textit{there
is no algorithm able to solve the problem for any alphabet}. As it will be
explained in details in the following sections, the main idea of the present
manuscript is to define a suitable heuristic \textit{in the space of all
alphabets} based on the following intuition. \textit{In the space of all
alphabets} there exist a suitable boundary separating the alphabets for which
the decision problem can be solved from the ones for which it cannot. Such
boundary is closely related to the \textit{transition to chaos in discrete
dynamical systems}. Based on this intuition, we propose very effective
heuristics (inspired by statistical mechanics). Such heuristics are able to
successfully\footnote{Despite the fact that, in principle, it is possible to
construct alphabets which satisfy our criteria until very large but finite
sizes of $n\times n$ squares and which fail to tile the whole plane
nevertheless.} identify all the known good tilings available in the literature
so far. To the best of authors knowledge, there are no other heuristics with
this property.

On the other hand, when a given alphabet does not satisfy such criteria (to be
introduced in the next section), the behavior of this alphabet tends to be (in
a precise sense to be specified in the next sections) chaotic when one
increases the size of the square to tile (something which prevents one from
guessing what happens in the case of very large squares). Thus, the relation
with chaotic discrete mappings suggests that alphabets which do not satisfy
our criterion could belong to the\textit{ region of the alphabets space} where
no effective heuristic to guess whether or not the given alphabet tiles the
plane is available.

Let's go back to the Wang tilings. Berger results were later improved by
Robinson \cite{[3]} which constructed a smaller aperiodic tile sets. A very
important work has been the one of Jeandel and Rao \cite{[4]} who showed that
the smallest aperiodic "tileset" must have eleven tiles (also providing
examples). A very powerful structure appearing in most of the known examples
of aperiodic tile sets is a hierarchical self-similar structure (notable
examples are \cite{[4]} and \cite{[4.1]}). However, there is one important
exception: the tiling discovered in \cite{[4.2]} which is neither periodic nor
self-similar (unlike most of the known aperiodic tilings).

The goal of the present paper is to define suitable temperature, entropy and
partition function associated to any alphabet $\Gamma$ in order to identify a
subset of alphabets which are good candidates to cover up the whole plane.
Thus, in order to achieve this partial decidability protocol (of an otherwise
undecidable problem), statistical mechanics plays a fundamental role.

The Wang tiling problem has further direct connections with statistical
physics. The aperiodic structures (which, on the mathematical side, are at the
heart of the undecidability of the Wang tiling problem) have been discovered
experimentally in \cite{[5]} \cite{[6]}. Interesting statistical mechanics
models (where the tiles represent clusters of atoms while the kinematical
constraints which these atoms must satisfy are represented as forbidden
matching between the tiles) took the Wang tiles as inspiration \cite{[7]}. In
these statistical models the above mentioned aperiodic structures play a
fundamental role. For instance, in \cite{[8.1]} (in a model with short range
interactions), it was discussed the appearance (in the self-similar structure
characteristic of Wang tiles) of an ordered phase with periodic behaviors at
increasingly larger scales as the temperature decreases. Closely related
examples of models with chaotic temperature dependence near zero (when the
Gibbs measure may diverge with, correspondingly, volatile variations in the
macroscopic state of the systems under investigation) have been constructed in
\cite{[9]} \cite{[10]} (for a very nice pedagogical review of these models see
\cite{[11]}\ and references therein).

The relations of the Wang tiling problem with physics are even deeper. These
can be seen in the analysis of undecidability in mathematics and physics
(which started soon after Godel results: see \cite{[11.1]} and references
therein). In recent years, we have observed a huge burst of interest in the
appearance of undecidability in fundamental physical problems. Two results
have influenced a lot the present work: the undecidability of the spectral gap
\cite{[13]}\ and the undecidability of thermalization in quantum many-body
\cite{[14]}. As far as the undecidability of the spectral gap, it has been
shown that, for 2D quantum many-body systems with nearest neighbor and
translational invariant interactions that the spectral gap is undecidable. As
far as the problem in thermalization is concerned, it has been shown that the
central issue to understand whether or not a given system thermalizes is
undecidable as well.

In the present manuscript we will exploit the relations between Wang tilings
and the appearance of chaotic structures. However, the present approach (as it
will be discussed in details in the next sections) is completely different
from the references described here above. The typical chaotic behavior which
appears in those references is closely related to the presence of aperiodic
tilings. On the other hand, in the present manuscript, we will associate a
suitable discrete mapping to any family of Wang tiles (namely, to any
\textit{Wang alphabet }$\Gamma$). When such mapping is not chaotic for a given
alphabet, then such an alphabet (which will be denoted as \textit{good
alphabet}) is a good candidate to cover up the whole plane. On the other hand,
when such mapping is chaotic, we cannot decide about the corresponding
alphabet. In other words, we will define \textit{suitable heuristics for
decidability protocol for Wang tilings}.

The connection between the present approach to the Wang tiling problem and the
physics of chaotic systems (for a review see \cite{[15]}) has its roots in the
linkage between deterministic chaos and undecidability which is already well
known since, at least, Chaitin's theorems \cite{15a0}. It is, by now, pretty
clear that determinism and undecidability can coexist and, therefore,
randomness and determinism can live together as well (see, for instance, the
detailed analysis in \cite{15a} \cite{15d} \cite{15e}). Very roughly speaking,
undecidability can manifest itself in dynamical systems with chaotic behavior
(in the present case, we will be interested in discrete dynamical systems). In
some cases, the existence of deterministic chaos in dynamical systems can be
related to a sort of Godelian undecidability rather than to the typical
numerical "intractability" (so characteristic of chaotic systems). Quite
fittingly, the discrete mapping that will be defined in the following sections
to identify good alphabets, manifests a clear chaotic tendency when the
alphabet is not good. Thus, the transition between good and bad alphabets
(within the space of all alphabets) looks similar to a transition from regular
to chaotic behavior in a discrete dynamical system.

\textbf{ }From the computer science viewpoint, the present results are also
very useful. The problem of determining whether a given finite set of square
tiles can tile a simply connected finite region of the plane is known to be NP-complete, a
complexity classification that implies no known polynomial-time algorithm
exists to solve it in the general case \cite{[22]}. This inherent
computational hardness means that exhaustive search methods, which require
exponential time and resources, are often infeasible for practical purposes.
Results such as\footnote{In \cite{[20]} the authors demonstrated that tiling
simply connected regions with a fixed set of rectangles is NP-complete, even
when the number of tile types is large but finite.} \cite{[20]} highlight the
necessity of developing efficient criteria to find good candidates to tile the
whole plane without resorting to exhaustive search. Hence, it becomes crucial
to develop criteria or "heuristics" that can efficiently suggest whether a
given alphabet $\Gamma$ is a good candidate or not without resorting to
brute-force search, which is prohibitive in terms of both memory and energy
consumption. Such recipes aim to identify polynomial-time verifiable
conditions or invariants that guarantee "tilability" or "non-tilability",
thereby circumventing the exponential complexity of the general problem. This
approach aligns with recent advances exemplified by \cite{[21]}, who provided
polynomial-time algorithms for special cases of tiling Polyominoes with
fixed-size square tiles and packing dominoes, leveraging compact geometric
representations of the regions. It is worth emphasizing that our approach can
be easily extended to different types of tiling problems (even in higher dimensions).

Our work builds upon these foundations by proposing a generalized framework
that defines combinatorial and geometric invariants serving as certificates of
"tilability" for tiles alphabets. By verifying these invariants, one can
preemptively determine the feasibility of tiling a region, thus avoiding
unnecessary computational overhead. This not only contributes to the
theoretical understanding of "tilability" within the P versus NP landscape but
also offers practical benefits in applications where resource constraints are critical.

This paper is organized as follows: in the second section we introduce the
discrete mapping associated to the Wang tiling problem together with the
"decidability" protocols. In the third section we discuss how to separate good
from bad alphabets. In the fourth section, we will discuss the relations of
the present approach with discrete chaos. In the fifth section, we will
discuss how to quantify "goodness" and "badness" of alphabets. In the sixth
section, we discuss what we have done and, especially, what we have not done.
In the final section some conclusions will be drawn.

\section{The Wang tiling problems and the $W_{\Gamma}(n)$ mapping}

As it has been already emphasized in the introduction, a Wang tiles set
$\Gamma$ or "alphabet" is a collections of a fixed number of squares (the
number of square tiles of the alphabet will be denoted as the size of the
alphabet $q=\left\vert \Gamma\right\vert $). Each edge of the square possesses
a color (a discrete label). The matching rules are the following:

\textbf{1) }The squares can be neither rotated nor reflected

\textbf{2) }Two square can be matched side to side only if the right edge of
the left square has the same color as the left edge of the right square.

\textbf{3) }Two square can be matched top to bottom only if the bottom edge of
the square on the top has the same color as the top edge of the square on the bottom.

Then, we want to know if the given alphabet $\Gamma$ of size $q$ can cover up
the whole plane. This question has been shown to be undecidable \textit{in the
general case} \cite{[2]}. However, it is extremely interesting to understand
how and when undecidability sets in in practice. The idea is to find a
concrete boundary in the space of all alphabets $\Gamma$ separating the
decidable instances from the undecidable ones. The results of the present
manuscript strongly suggest that in the space of all alphabets such a boundary
can be defined and, moreover, share many features with the transition to chaos
in discrete dynamical systems and can be identified using sound arguments from
statistical mechanics.

In order to proceed, let us define the following quantity:
\begin{equation}
W_{\Gamma}\left(  n\right)  \overset{def}{=}\ \left\{
number\ of\ different\ tilings\ of\ a\ n\times
n\ squares\ with\ the\ alphabet\ \Gamma\right\}  \ . \label{map1}%
\end{equation}
Let us also define \textit{the energy} $E_{\Gamma}(n)$ \textit{of an alphabet}
$\Gamma$ on a $n\times n\ square$ as the area $A(n)$ of the square (namely
$n^{2}$):%
\begin{equation}
E_{\Gamma}(n)\overset{def}{=}A(n)=n^{2}\ . \label{map1.5}%
\end{equation}
Thus, from the above definitions it is clear that $W_{\Gamma}\left(  n\right)
$ \textit{plays the role of the degeneracy of the energy level} $E_{\Gamma
}(n)$. Therefore, we can define \textit{the entropy} $S_{\Gamma}(n)$ of
$\Gamma$ \textit{at the energy level} $n$ as:
\begin{equation}
S_{\Gamma}(n)\overset{def}{=}\log W_{\Gamma}\left(  n\right)  \ .
\label{map1.75}%
\end{equation}
In the following it will be convenient to use the $\log_{10}$ instead of the
usual $\ln$, but conceptually there is no difference. These definitions are
natural for the following reasons.

\textit{On the one hand}, we expect that for alphabets $\Gamma$ with good
thermodynamical behavior there will be many different ways to tile a $n\times
n$ square (since for most of physically reasonable systems the degeneracy of
the energy level is an increasing function of the energy). This, in
particular, implies that $W_{\Gamma}\left(  n\right)  $ is different from zero
for arbitrarily large $n$. This last conclusion (which is a rather obvious
fact in many physical systems) precisely means that $\Gamma$ can tile
arbitrarily large regions! Hence, for such \textit{good alphabets} (namely,
alphabets with a good thermodynamical behavior), both $W_{\Gamma}\left(
n\right)  $ and $S_{\Gamma}(n)$ should be increasing function of $n$ and,
consequently, good alphabets can tile arbitrarily large regions. This last
observation is equivalent to the typical thermodynamical condition related to
the positiveness of the temperature:
\begin{equation}
\frac{\partial S}{\partial E}>0\Leftrightarrow\ \frac{\partial S_{\Gamma}%
(n)}{\partial n}>0\ . \label{map1.9}%
\end{equation}

\textit{Conversely}, from the intuitive viewpoint one can assume that an
alphabet $\Gamma$ with a good thermodynamical limit\footnote{Namely, an
alphabet which can tile aribitrarily large squares and which, therefore,
allows to define the thermodynamical limit with the above definitions of
entropy and energy in Eqs. (\ref{map1.5}) and (\ref{map1.75}).} should have a
well defined positive temperature and, consequently, Eq. (\ref{map1.9}) should
be satisfied.

A very important function which captures the main thermodynamical features of
a system is its partition function. Using the previous definitions we can
define \textit{the partition function of the alphabet} $\Gamma$ \textit{as}
$Z_{\Gamma}$:%
\begin{equation}
Z_{\Gamma}(\beta)\overset{def}{=}\sum_{n}W_{\Gamma}\left(  n\right)
\exp\left(  -\beta n^{2}\right)  \ . \label{map1.99}%
\end{equation}
The above partition function encodes relevant informations about the alphabet
$\Gamma$. For instance, if the alphabet $\Gamma$ does not tile the plane then
there exist an $n^{\ast}$ with the property that
\[
\forall n>n^{\ast}\ ,\ \ W_{\Gamma}\left(  n\right)  =0
\]
(as otherwise, $\Gamma$ would actually tile the plane\footnote{Indeed, the
negation of the formula "$\forall n>n^{\ast}\ ,\ \ W_{\Gamma}\left(  n\right)
=0$"\ is precisely that $W_{\Gamma}(n)$ is different from zero for arbitrarily
large $n$ and so $\Gamma$ tiles the plane.}). In this case, the partition
function is analytic as $Z_{\Gamma}(\beta)$ will be just a polynomial in
$\exp(-\beta)$. When $Z_{\Gamma}(\beta)$ is not analytic as function of
$\beta$ then necessarily there are infinitely many terms in the sum on the
right hand side of Eq. (\ref{map1.99}) (otherwise the function would be
analytic) and this means that actually $\Gamma$ tiles the whole plane.
Consequently, the properties of $Z_{\Gamma}(\beta)$ are also worth to be
further investigated (we will come back on them in a future publication), but
here below we will discuss a more direct approach to disclose the arising of undecidability.

\subsection{Protocol I}

In order to devise a partial "decidability" protocol able to detect a subclass
of good alphabet, the requirement to satisfy the condition in Eq.
(\ref{map1.9}) is a good starting point. Obviously, from the practical
viewpoint, one cannot compute $W_{\Gamma}\left(  n\right)  $ for arbitrarily
large $n$ due to the fact that this would need an infinitely powerful computer
(which is not available). Thus, in order to use the present ideas based on
statistical mechanics in concrete situations, one can only verify that the
condition in Eq. (\ref{map1.9}) is satisfied for finite $n$. Thus, the
question is: how large should the set ($1$, $2$, $3$,.., $n_{\max}$) be in
order to be confident that the numerical verification of Eq. (\ref{map1.9})
for ($1$, $2$, $3$,.., $n_{\max}$) ensures "tilability for arbitrarily large
$n$"? We do not have rigorous results on this point. However, the numerical
results to be described in the following sections strongly suggest that it
could be enough to verify Eq. (\ref{map1.9}) for all $n\leq n_{\max}$ with
\begin{equation}
n_{\max}\gg\left\vert \Gamma\right\vert =q. \label{protocol1a}%
\end{equation}
In the examples here below it has been enough to take $n_{\max}$ 2 or 3 times
$q$. This strategy is very simple to implement.

\subsection{Protocol II}

On the other hand, it is possible to define a second criterion to detect the
"goodness" of a given alphabet $\Gamma$ which is also able to disclose a very
intriguing relation with the transition to chaos in discrete dynamical
systems. The main idea is to plot $W_{\Gamma}\left(  n+1\right)  $ in terms of
$W_{\Gamma}\left(  n\right)  $. In this way, the mapping $f_{\Gamma}$ is
computed numerically plotting on the vertical $Y-$axis $W_{\Gamma}\left(
n+1\right)  $ and on the horizontal $X-$axis $W_{\Gamma}\left(  n\right)  $.
In this plots, the two coordinates of each point $P$ are:
\begin{equation}
P=(X,Y)=\left(  W_{\Gamma}\left(  n\right)  ,W_{\Gamma}\left(  n+1\right)
\right)  \ .\label{map2}%
\end{equation}
The mapping obtained in this way defines (at least locally) $W_{\Gamma}\left(
n+1\right)  $ as a function $F_{\Gamma}$ of $W_{\Gamma}\left(  n\right)  $:
\begin{equation}
W_{\Gamma}\left(  n+1\right)  =F_{\Gamma}\left(  W_{\Gamma}\left(  n\right)
\right)  \ .\label{map3}%
\end{equation}
Such function $F_{\Gamma}$ is the sought discrete mapping. We will do the same
with the entropy, plotting $S_{\Gamma}\left(  n+1\right)  $ in terms of
$S_{\Gamma}\left(  n\right)  $ obtaining
\begin{equation}
S_{\Gamma}\left(  n+1\right)  =f_{\Gamma}\left(  S_{\Gamma}\left(  n\right)
\right)  \ ,\label{map3.1}%
\end{equation}
$S_{\Gamma}\left(  n+1\right)  $ as a function $f_{\Gamma}$ of $S_{\Gamma
}\left(  n\right)  $. Due to the undecidable nature of this combinatorial
problem, it may be very difficult to be able to compute explicitly such
function $f_{\Gamma}$ for any alphabet $\Gamma$. However, as it will be now
discussed, this is not really necessary. The present analysis shows that
$f_{\Gamma}$ behaves very differently depending on whether or not there is an
"easy" algorithm to decide when $\Gamma$ is a good candidate to cover the
whole plane. These two very different behaviors allows to deduce many
non-trivial results even if the analytic form of $f_{\Gamma}$ is not known precisely.

\section{Separating "good" from "bad" mappings}

To properly assess and classify the different alphabets as either ``good'' or
``bad'', we analyze for each case three key visualizations: the geometric
structure of the tiling, the complexity growth plot $W_{\Gamma}(n)$ versus
$n$, and the phase-space plot $W_{\Gamma}(n+1)$ versus $W_{\Gamma}(n)$. These
plots jointly offer a comprehensive perspective on the spatial and
combinatorial dynamics associated with each alphabet. For each case, a single
figure presents the three components together, enabling a direct visual
comparison of their respective behaviors. In the discussion that follows, we
adopt the notation $G_{i}$ to denote alphabets that are considered ``Good''
either due to established results in the literature or based on favorable
tiling behavior observed through simulation. Conversely, alphabets denoted as
$B_{i}$ are classified as ``Bad'', reflecting either their known limitations
as reported in the literature or pathological behaviors detected in
computational experiments. This classification guides our comparative analysis
and is supported by the visual and quantitative characteristics displayed on
each plots. Further details will include presenting similar plots for the
associated entropy $S_{\Gamma}(n)$, which will offer additional insights into
the alphabet's thermodynamic behavior.

\begin{figure}[th]
\vspace{0.1em}  \centering
\begin{subfigure}{0.4\linewidth}
		\centering
		\includegraphics[width=0.5\linewidth]{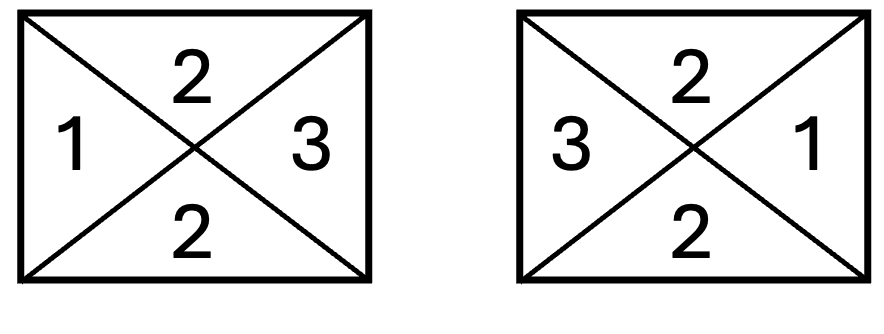}
		\caption{Alphabet G1}
		\label{fig:G1_analysis_a}
	\end{subfigure}
\hfill
\par
\vspace{0.5em}
\par
\begin{subfigure}{\linewidth}
		\centering
		\includegraphics[width=0.8\linewidth]{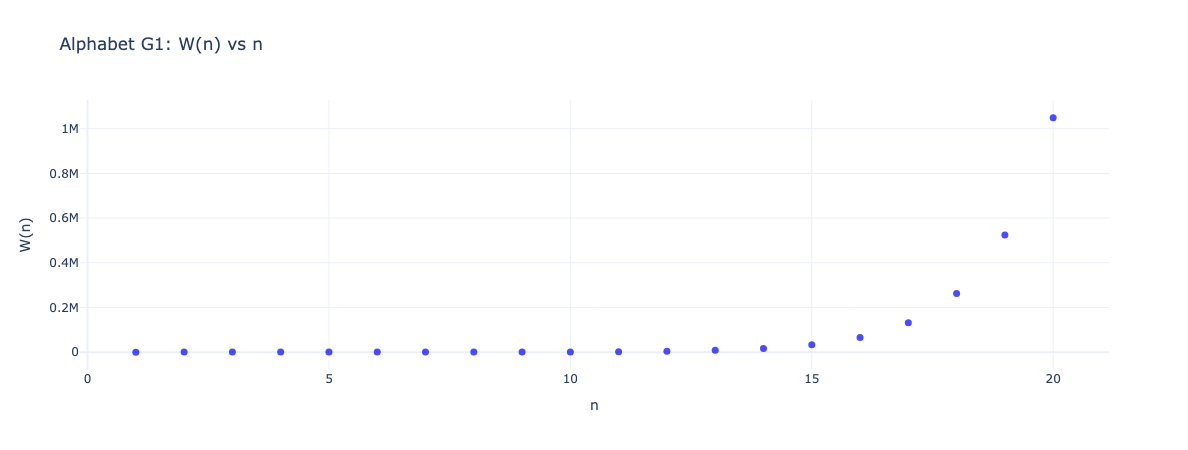}
		\caption{$W_{\Gamma}(n)$ vs $n$}
		\label{fig:G1_analysis_b}
	\end{subfigure}
\par
\vspace{0.5em}
\par
\begin{subfigure}{\linewidth}
		\centering
		\includegraphics[width=0.8\linewidth]{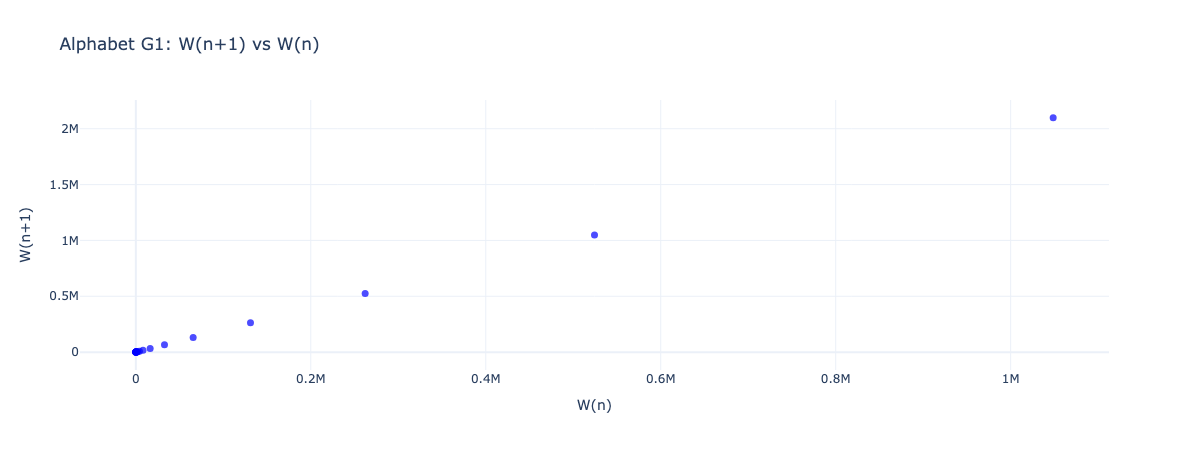}
		\caption{$W_{\Gamma}(n+1)$ vs $W_{\Gamma}(n)$}
		\label{fig:G1_analysis_c}
	\end{subfigure}
\caption{Alphabet G1 and its associated complexity plots.}%
\label{fig:G1_analysis}%
\end{figure}

\textbf{Alphabet G1.} This is a straightforward minimal example of a
\textquotedblleft good\textquotedblright\ alphabet. It consists of only two
tiles, yet it is capable of completely mapping the two-dimensional plane. The
plot of $W_{\Gamma}(n)$ depicts a simple alphabet exhibiting monotonic growth
in all directions. Consequently, $W_{\Gamma}(n)$ increases with structure size
$n$ , and the corresponding $W_{\Gamma}(n+1)$ vs. $W_{\Gamma}(n)$ plot exhibit
linear growth , reflecting the inherent simplicity and full coverage of the
alphabet. In particular, $W_{\Gamma}(n)=2^{n}$ (as any sequence of colors 1
and 3 on the left boundary of the square provides a unique tiling\footnote{We
thank the referee for this remark.}). Illustrations can be seen on
\textbf{Fig.}~\ref{fig:G1_analysis}.

\begin{figure}[th]
\centering
\par
\begin{subfigure}{0.7\linewidth}
		\centering
		\includegraphics[width=0.5\linewidth]{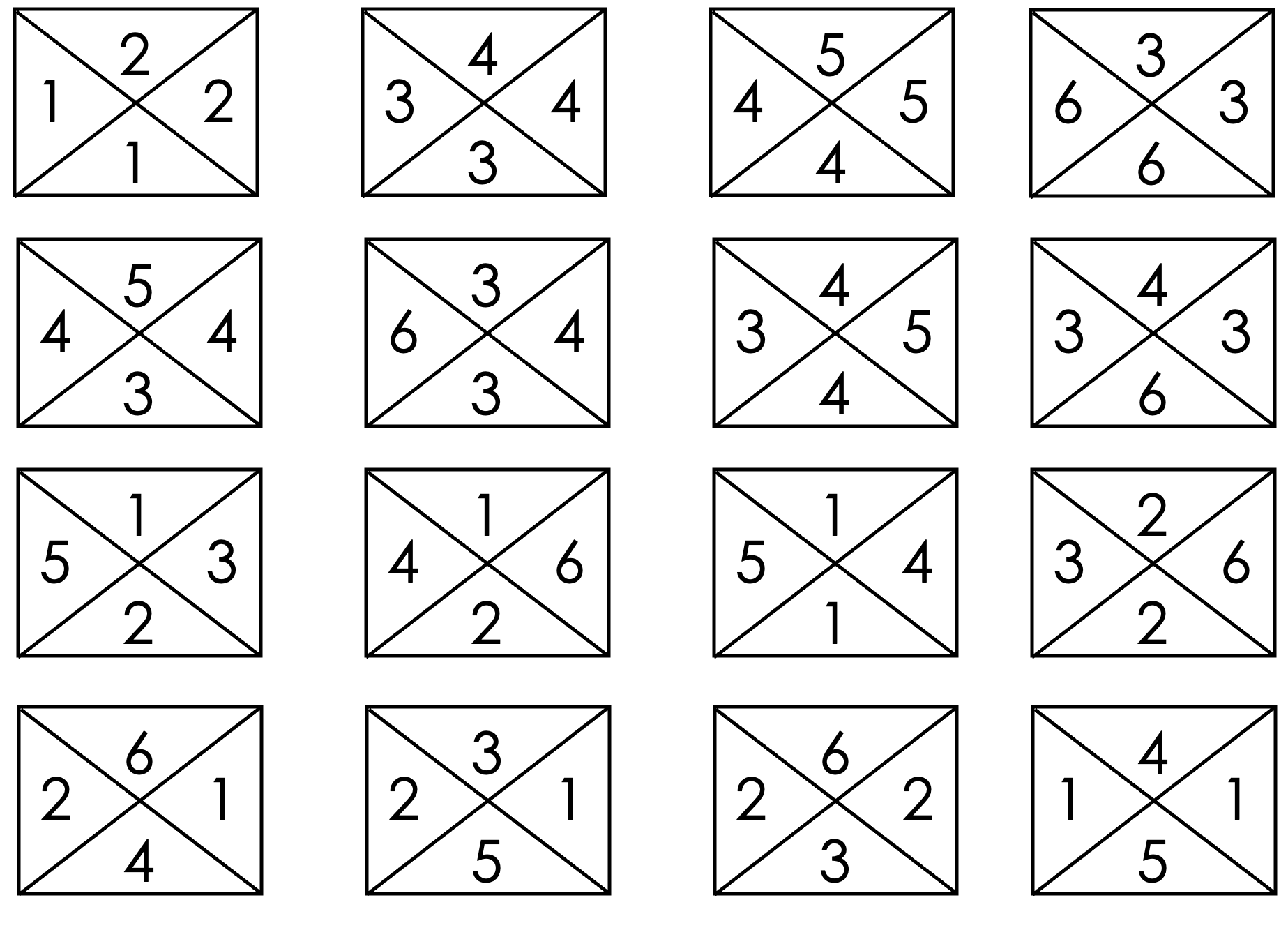}
		\caption{Alphabet G2}
		\label{fig:G2_analysis_a}
	\end{subfigure}
\hfill
\par
\vspace{0.5em}  \begin{subfigure}{\linewidth}
		\centering
		\includegraphics[width=0.8\linewidth]{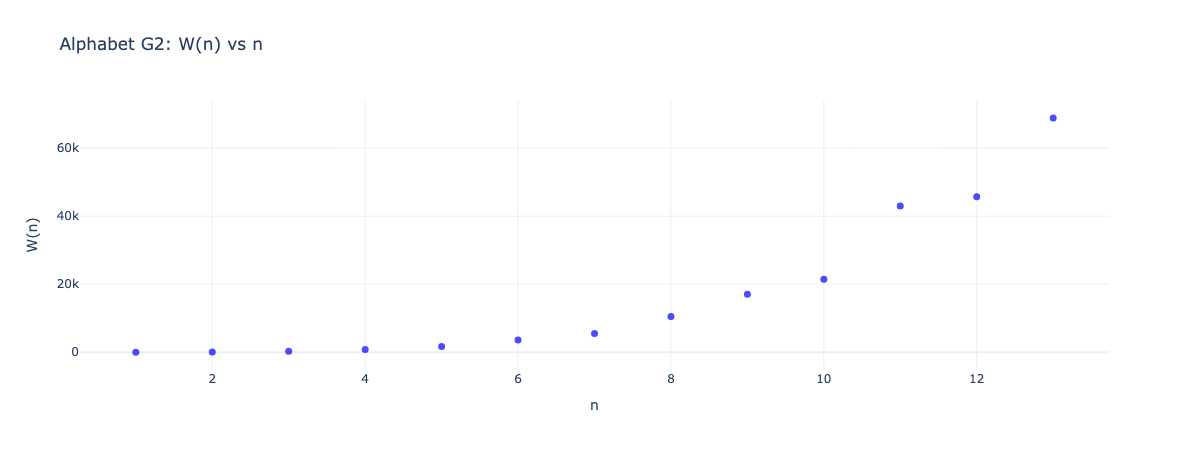}
		\caption{$W_{\Gamma}(n)$ vs $n$}
		\label{fig:G2_analysis_b}
	\end{subfigure}
\par
\vspace{0.5em}
\par
\begin{subfigure}{\linewidth}
		\centering
		\includegraphics[width=0.8\linewidth]{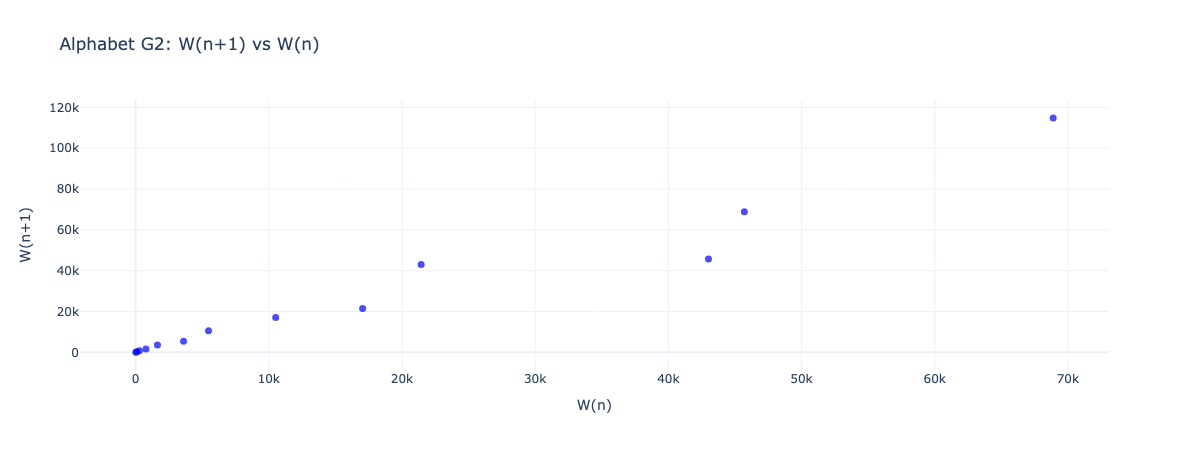}
		\caption{$W_{\Gamma}(n+1)$ vs $W_{\Gamma}(n)$}
		\label{fig:G2_analysis_c}
	\end{subfigure}
\caption{Alphabet G2 and its associated complexity plots.}%
\label{fig:G2_analysis}%
\end{figure}

\textbf{Alphabet G2.} Cited from \cite{[19]}, this alphabet is an example of
an aperiodic, self-similar tiling system. The behavior of $W_{\Gamma}(n)$ is
nonlinear but exhibits a recognizable monotone behavior. The $W_{\Gamma}(n+1)$
vs. $W_{\Gamma}(n)$ plot reveals the recursive and hierarchical construction
of the tiling. This self-similarity is a distinguishing feature that supports
its classification as a \textquotedblleft good\textquotedblright\ alphabet.
Illustrations are provided in \textbf{FIG.}~\ref{fig:G2_analysis}.

\begin{figure}[th]
\centering
\par
\begin{subfigure}{0.7\linewidth}
		\centering
		
		\includegraphics[width=0.5\linewidth]{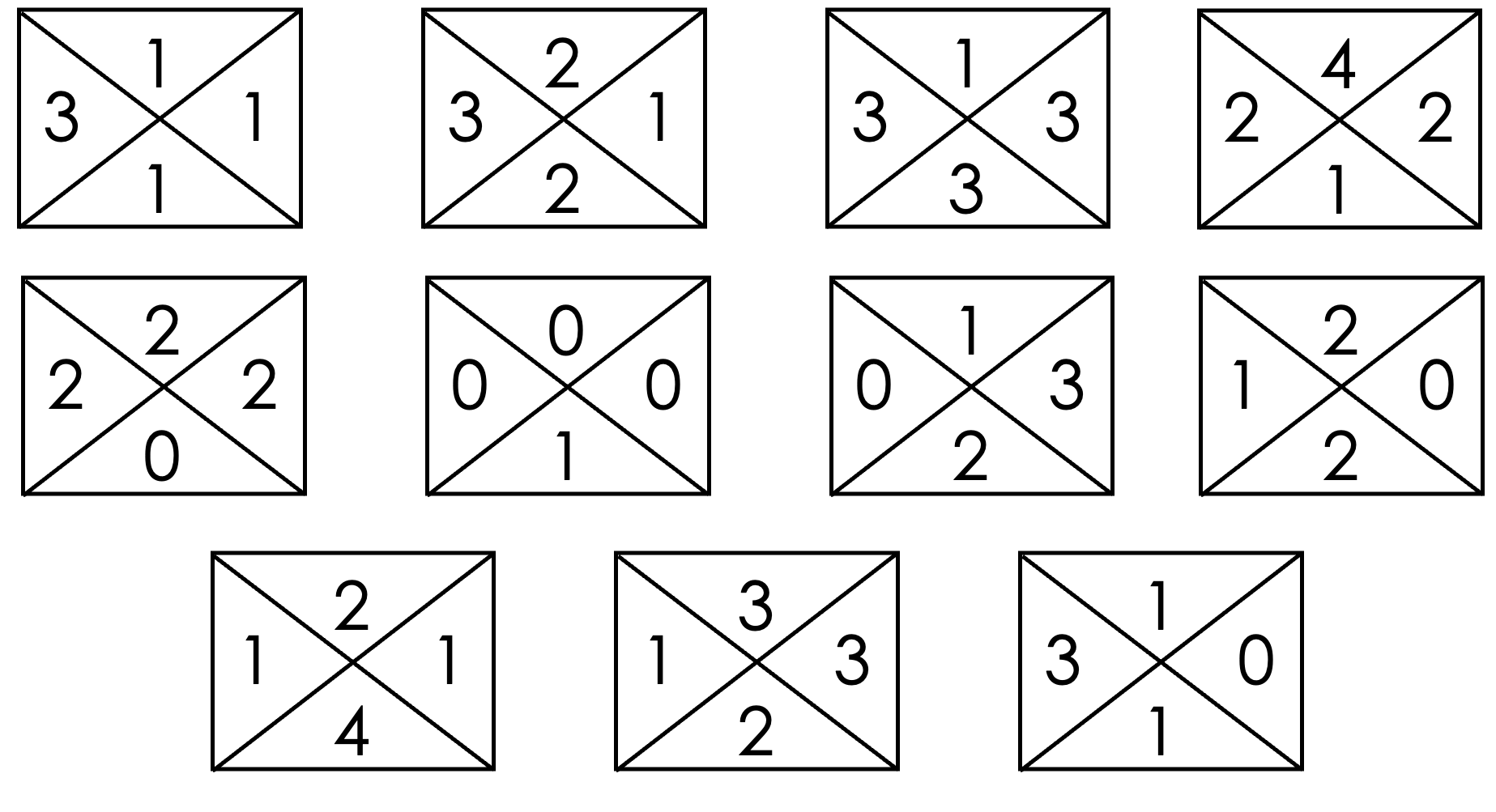}
		\caption{Alphabet G3}
		\label{fig:G3_analysis_a}
	\end{subfigure}
\par
\vspace{0.8em}
\par
\begin{subfigure}{\linewidth}
		\centering
		\includegraphics[width=0.8\linewidth]{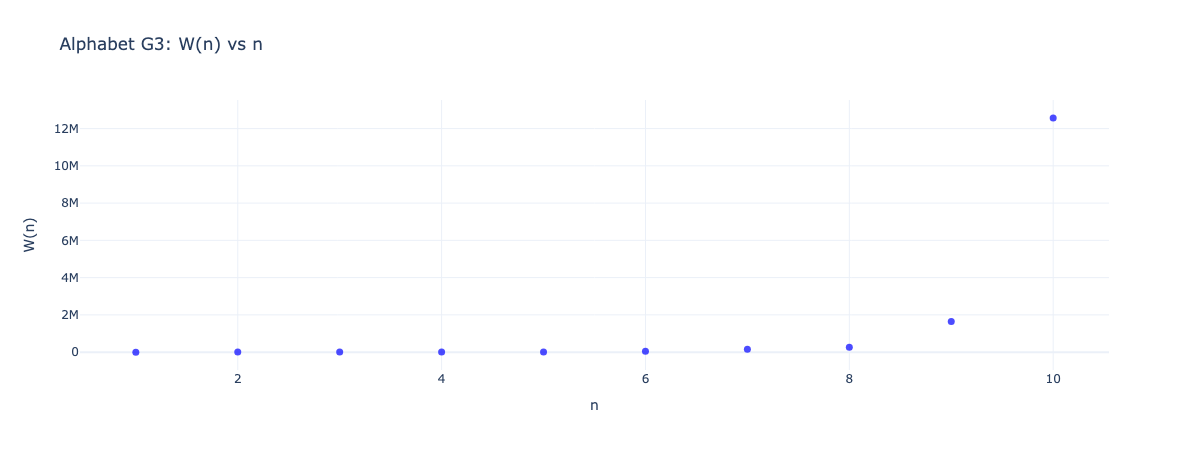}
		\caption{$W_{\Gamma}(n)$ vs $n$}
		\label{fig:G3_analysis_b}
	\end{subfigure}
\par
\vspace{0.8em}
\par
\begin{subfigure}{\linewidth}
		\centering
		\includegraphics[width=0.8\linewidth]{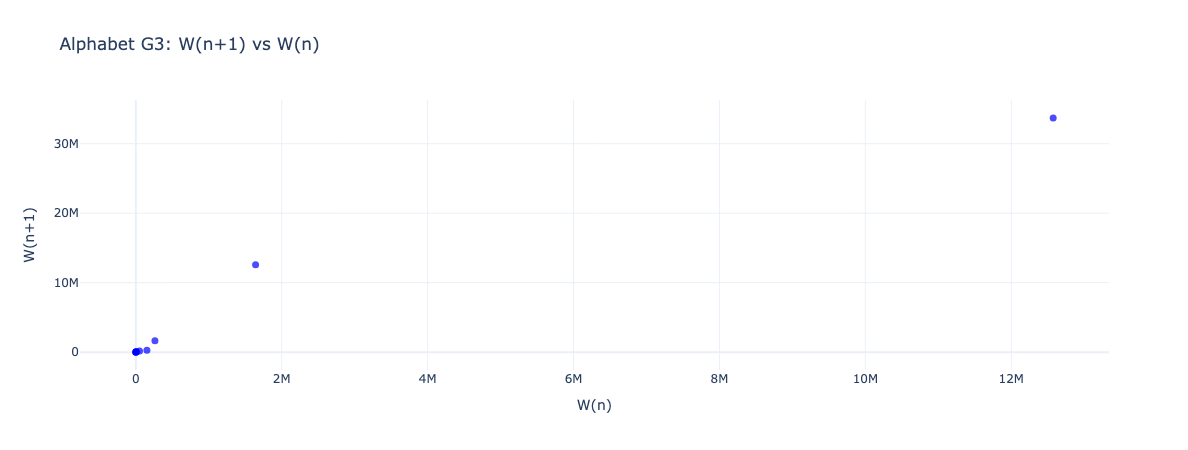}
		\caption{$W_{\Gamma}(n+1)$ vs $W_{\Gamma}(n)$}
		\label{fig:G3_analysis_c}
	\end{subfigure}
\caption{Alphabet G3 and its associated complexity plots.}%
\label{fig:G3_analysis}%
\end{figure}

\textbf{Alphabet G3.} Described in \cite{[4]}, this alphabet represents an
aperiodic but non-self-similar system. Unlike G2, the lack of self-similarity
manifests in more irregularities within the $W_{\Gamma}(n)$ growth, and the
$W_{\Gamma}(n+1)$ vs. $W_{\Gamma}(n)$ plot does not present a fractal pattern.
However, the absence of periodicity and the tiling completeness render it a
member of the \textquotedblleft good\textquotedblright\ class, albeit with
more complex dynamical behavior. Illustrations can be seen on \textbf{FIG.}%
~(\ref{fig:G3_analysis})

\begin{figure}[th]
\centering
\par
\begin{subfigure}{0.7\linewidth}
		\centering
		\includegraphics[width=0.5\linewidth]{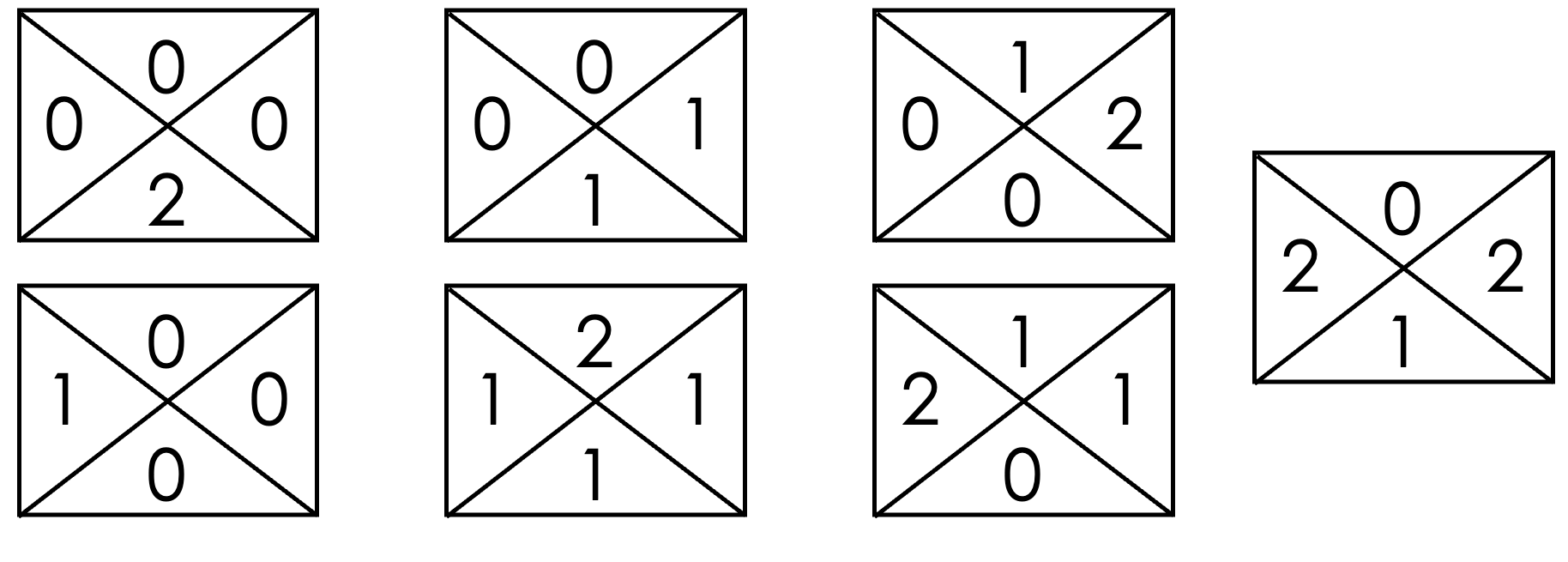}
		\caption{Alphabet B1}
		\label{fig:B1_analysis_a}
	\end{subfigure}
\par
\vspace{0.8em}
\par
\begin{subfigure}{\linewidth}
		\centering
		\includegraphics[width=0.8\linewidth]{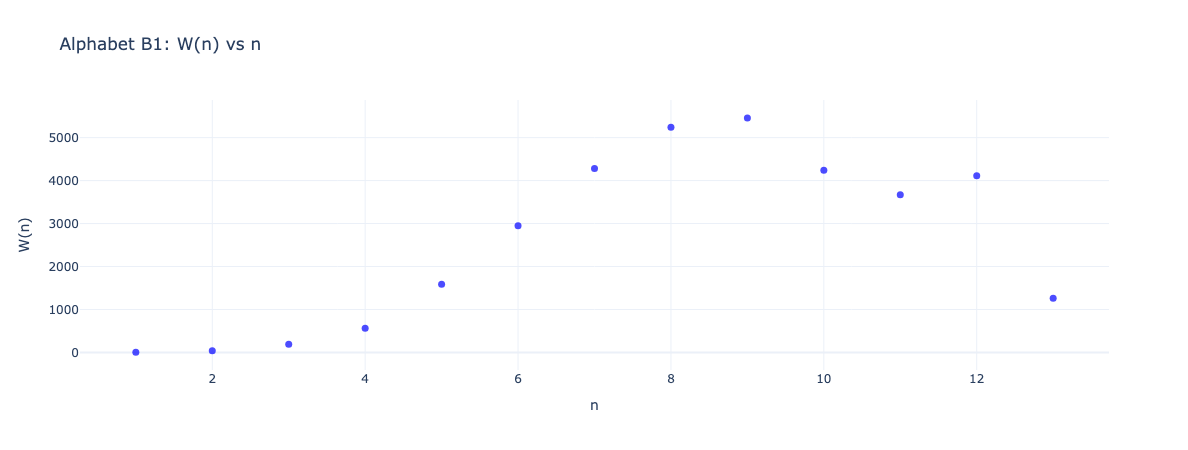}
		\caption{$W_{\Gamma}(n)$ vs $n$}
		\label{fig:B1_analysis_b}
	\end{subfigure}
\par
\vspace{0.8em}
\par
\begin{subfigure}{\linewidth}
		\centering
		\includegraphics[width=0.8\linewidth]{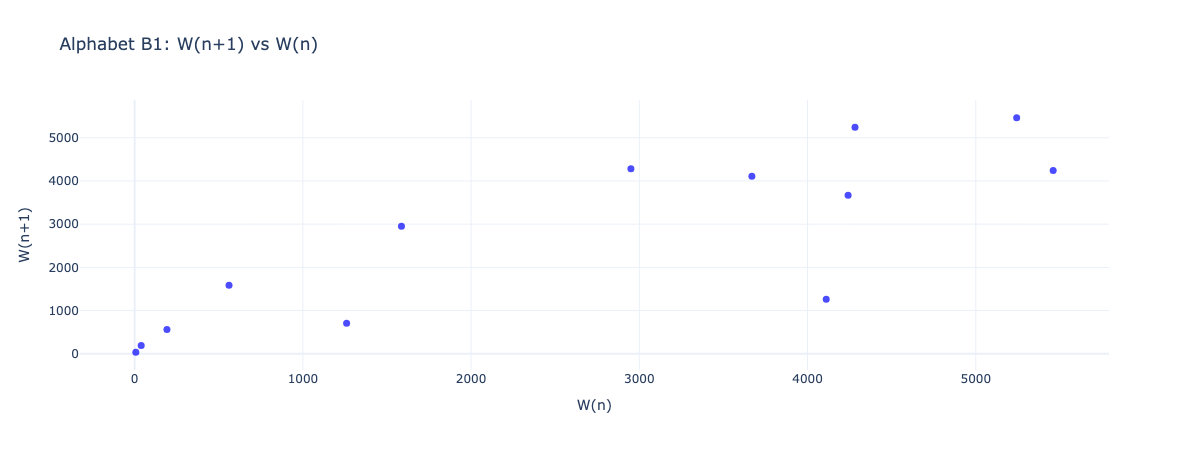}
		\caption{$W_{\Gamma}(n+1)$ vs $W_{\Gamma}(n)$}
		\label{fig:B1_analysis_c}
	\end{subfigure}
\caption{Alphabet B1 and its associated complexity plots.}%
\label{fig:B1_analysis}%
\end{figure}

\textbf{Alphabet B1.} This simulated \textquotedblleft bad\textquotedblright%
\ alphabet demonstrates irregular growth in $W_{\Gamma}(n)$ and exhibits
discontinuous transitions in the $W_{\Gamma}(n+1)$ vs. $W_{\Gamma}(n)$ plot.
The lack of structure in both plots reflects the limitations of the alphabet
to generate a coherent or comprehensive coverage of the plane. Illustrations
can be seen on \textbf{FIG.}~(\ref{fig:B1_analysis})

\begin{figure}[th]
\centering
\par
\begin{subfigure}{0.7\linewidth}
		\centering
		\includegraphics[width=0.5\linewidth]{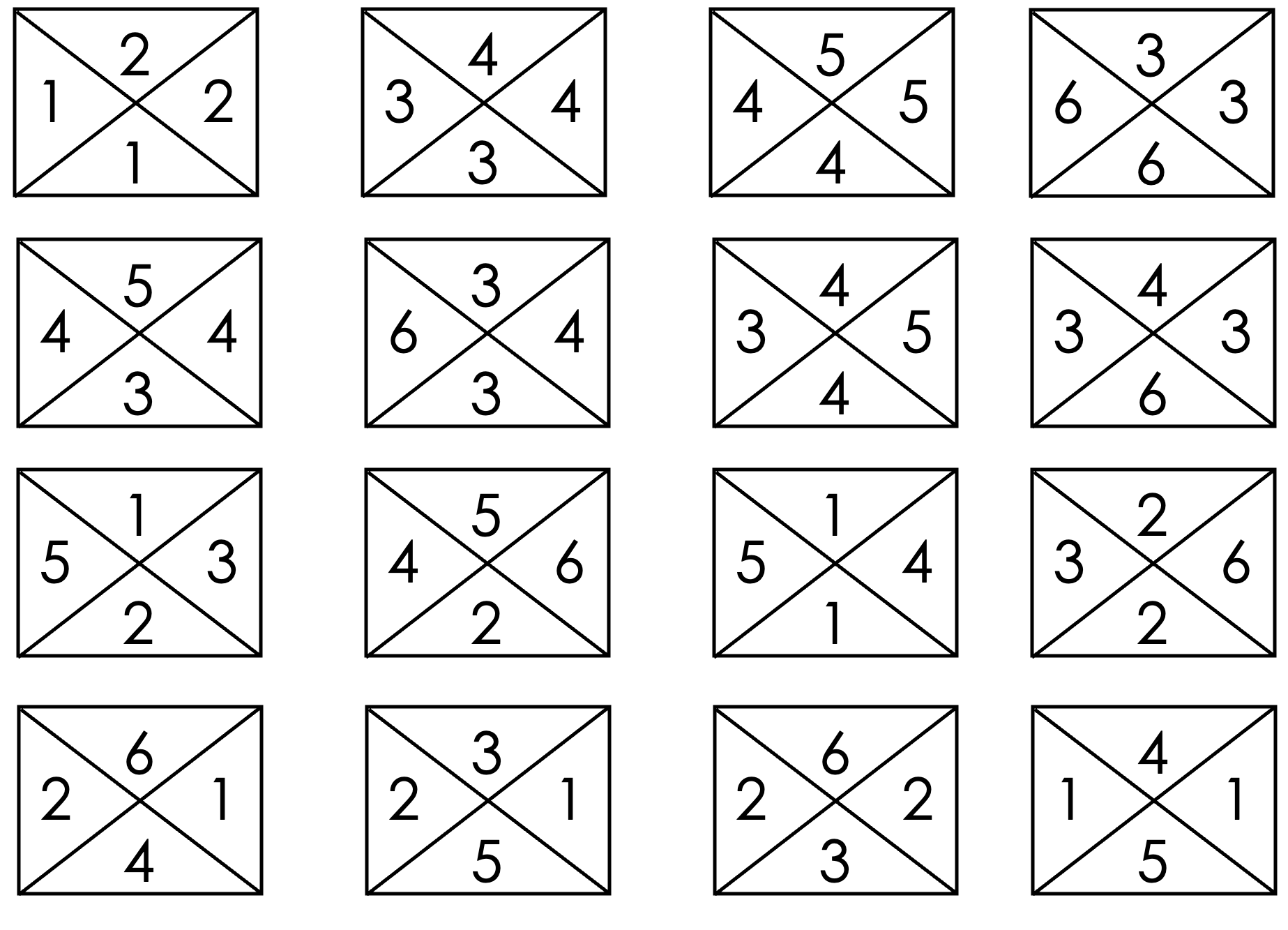}
		\caption{Alphabet B2}
		\label{fig:B2_analysis_a}
	\end{subfigure}
\par
\vspace{0.8em}
\par
\begin{subfigure}{\linewidth}
		\centering
		\includegraphics[width=0.5\linewidth]{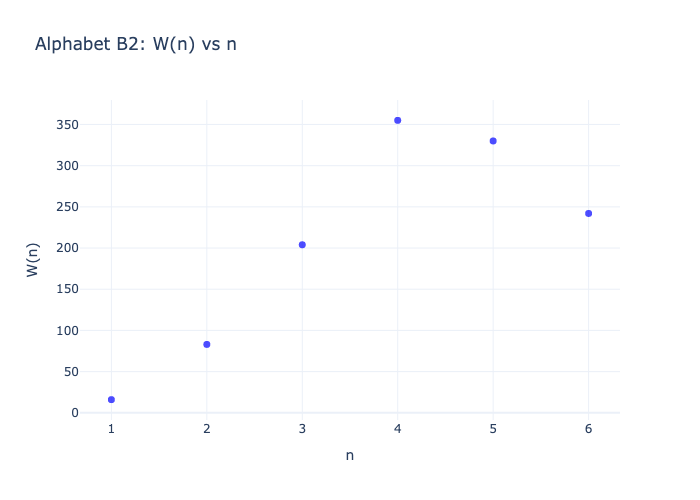}
		\caption{$W_{\Gamma}(n)$ vs $n$}
		\label{fig:B2_analysis_b}
	\end{subfigure}
\par
\vspace{0.8em}
\par
\begin{subfigure}{\linewidth}
		\centering
		\includegraphics[width=0.5\linewidth]{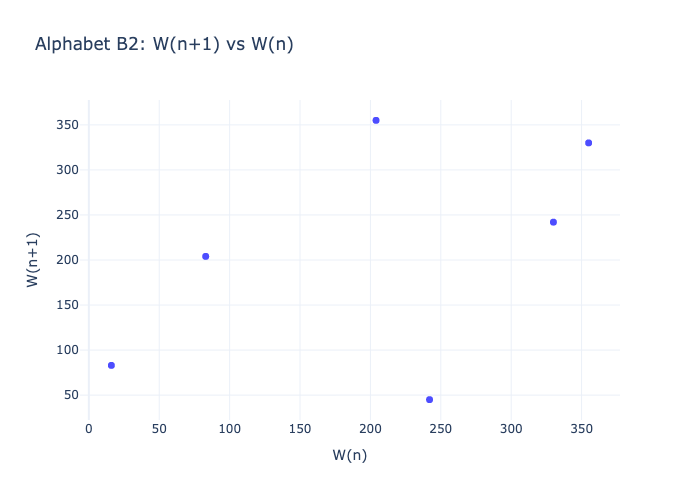}
		\caption{$W_{\Gamma}(n+1)$ vs $W_{\Gamma}(n)$}
		\label{fig:B2_analysis_c}
	\end{subfigure}
\caption{Alphabet B2 and its associated complexity plots.}%
\label{fig:B2_analysis}%
\end{figure}

\textbf{Alphabet B2.} This alphabet highlights the system's sensitivity to
minor tile modifications. It is derived directly from the ``good'' alphabet G2
by altering a single color on one tile: tile 10, which was changed from (1, 6,
2, 4) in G2 to (5, 6, 2, 4) for this set. This single change is sufficient to
induce \textquotedblleft bad\textquotedblright\ behavior. As illustrated in
\textbf{FIG.}~(\ref{fig:B2_analysis}), the alphabet fails to produce a stable
or predictable $W_{\Gamma}(n)$ trajectory. The significant scatter in the
$W_{\Gamma}(n+1)$ vs. $W_{\Gamma}(n)$ plot further indicates poor local
consistency and an absence of long-range ordering, reinforcing this categorization.

\begin{figure}[th]
\centering
\par
\begin{subfigure}{0.7\linewidth}
		\centering
		\includegraphics[width=0.5\linewidth]{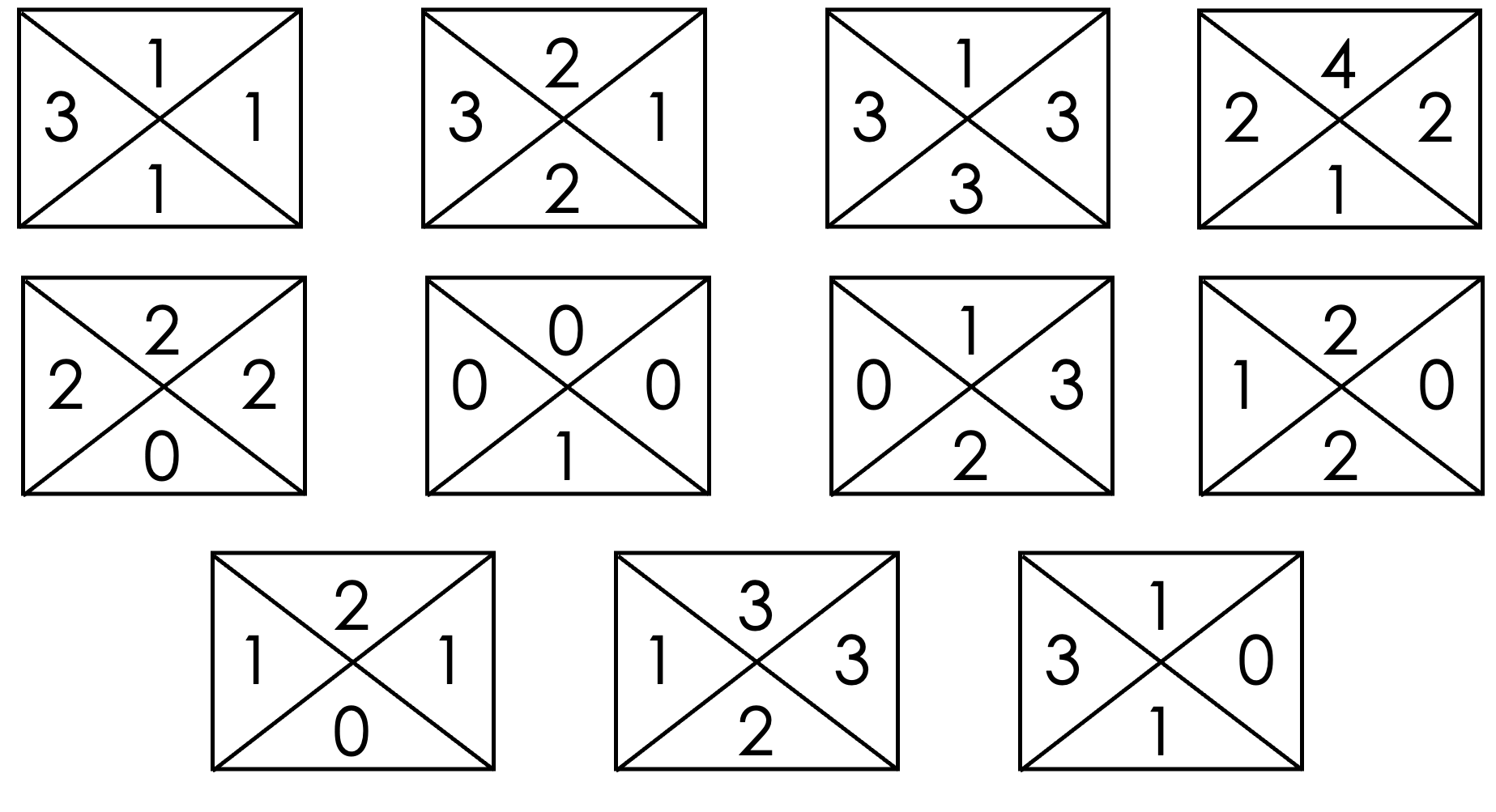}
		\caption{Alphabet B3}
		\label{fig:B3_analysis_a}
	\end{subfigure}
\par
\vspace{0.8em}
\par
\begin{subfigure}{\linewidth}
		\centering
		\includegraphics[width=0.8\linewidth]{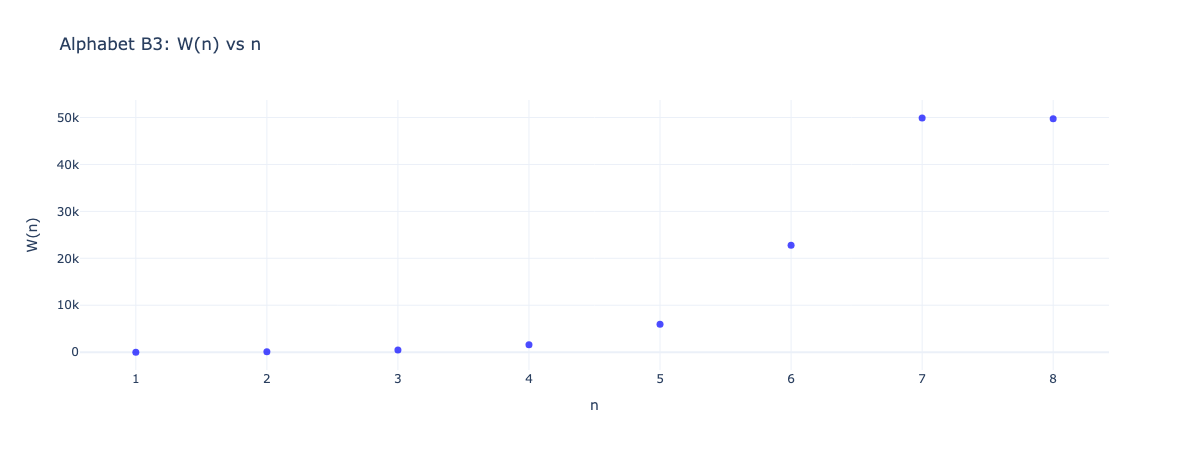}
		\caption{$W_{\Gamma}(n)$ vs $n$}
		\label{fig:B3_analysis_b}
	\end{subfigure}
\par
\vspace{0.8em}
\par
\begin{subfigure}{\linewidth}
		\centering
		\includegraphics[width=0.8\linewidth]{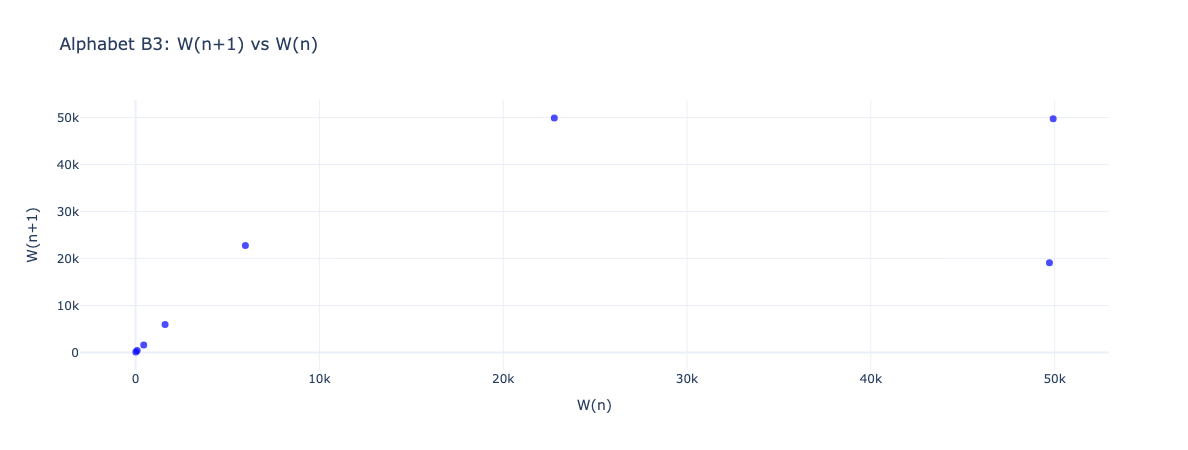}
		\caption{$W_{\Gamma}(n+1)$ vs $W_{\Gamma}(n)$}
		\label{fig:B3_analysis_c}
	\end{subfigure}
\caption{Alphabet B3 and its associated complexity plots.}%
\label{fig:B3_analysis}%
\end{figure}

\textbf{Alphabet B3.} Similar to the B2 case, this alphabet demonstrates the
system's sensitivity to small changes. Its anomalous behavior originates from
a single color modification in tile 9, which was changed from (2, 1, 4, 1) to
(2, 1, 0, 1). This alphabet displays anomalous behavior in both plots.
Although its $W_{\Gamma}(n)$ initial growth superficially resembles that of
some well-behaved alphabets, its $W_{\Gamma}(n+1)$ vs. $W_{\Gamma}(n)$ plot
uniquely reveals a stagnant growth that leads to structural inconsistencies.
This significant deviation from expected patterns suggests a transitional or
deceptive character, challenging straightforward classification. Illustrations
can be seen in \textbf{FIG.}~(\ref{fig:B3_analysis}).

\medskip

Through this comparative study, it becomes clear that the distinction between
``good'' and ``bad'' alphabets is more nuanced than a binary classification.
The behavior of $W_{\Gamma}$ under sequential growth, as visualized in the two
standard plots, offers a powerful lens to probe underlying structural
properties and dynamic complexity of tiling alphabets.

To effectively compare alphabets with vastly different $W_{\Gamma}(n)$ values,
it is crucial to address the scale issue. Directly analyzing $W_{\Gamma}(n)$
can obscure underlying patterns, as its magnitude can vary widely across
different alphabets. Therefore, we propose utilizing the entropy $S_{\Gamma
}(n)$ as the primary variable for analysis, defined as:%

\begin{equation}
S_{\Gamma}(n)=\log_{10}(W_{\Gamma}(n)) \label{log}%
\end{equation}

This logarithmic transformation effectively normalizes the scale of
$W_{\Gamma}(n)$ across diverse alphabets, thereby enabling a more meaningful
and robust comparison, particularly within the context of regression tasks.
Such normalization is essential for achieving consistent and interpretable
analytical results. The significant impact of this transformation on
regression analysis is clearly demonstrated in \textbf{Table
\ref{regression_results}}, which presents the derived regression coefficients
using $S_{\Gamma}(n+1)$ vs $S_{\Gamma}(n)$.

\begin{figure}[th]
\centering
\par
\begin{subfigure}{0.5\linewidth}
		\centering
		\includegraphics[width=0.5\linewidth]{figures/Alphabet_G1.png}
		\caption{Alphabet G1}
		\label{fig:G1_Reg_a}
	\end{subfigure}
\par
\vspace{0.8em}
\par
\begin{subfigure}{\linewidth}
		\centering
		\includegraphics[width=0.8\linewidth]{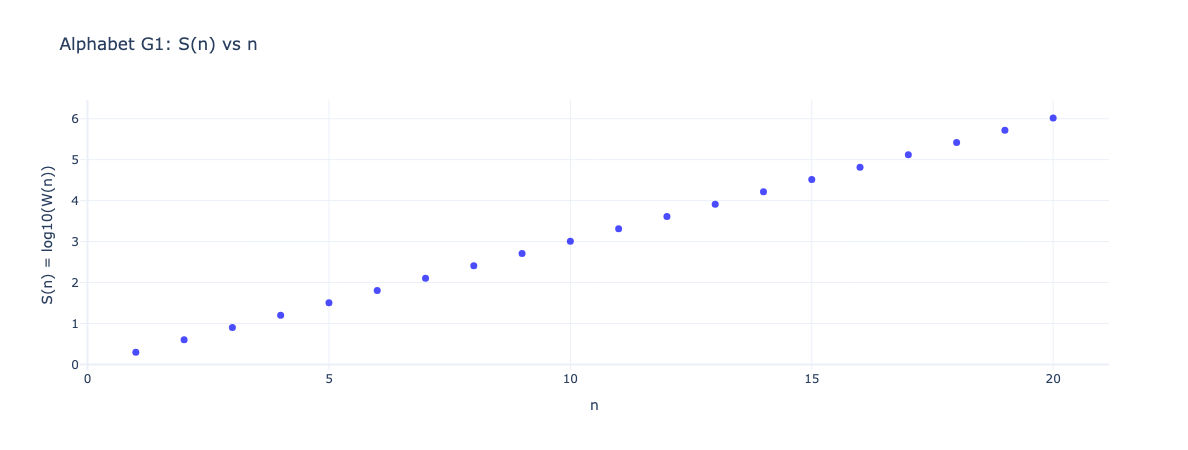}
		\caption{$S_{\Gamma}(n)$ vs $n$}
		\label{fig:G1_Reg_b}
	\end{subfigure}
\par
\vspace{0.8em}
\par
\begin{subfigure}{\linewidth}
		\centering
		\includegraphics[width=0.8\linewidth]{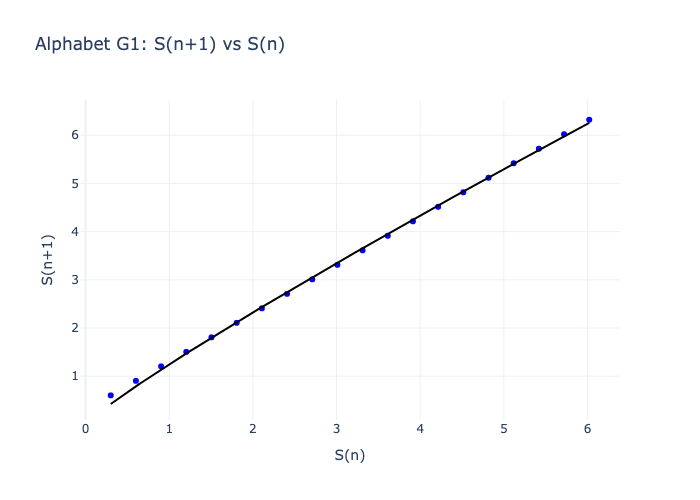}
		\caption{$S_{\Gamma}(n+1)$ vs $S_{\Gamma}(n)$}
		\label{fig:G1_Reg_c}
	\end{subfigure}
\caption{Alphabet G1 and its associated entropy plots.}%
\label{fig:G1_Reg}%
\end{figure}

\begin{figure}[th]
\centering
\par
\begin{subfigure}{0.7\linewidth}
		\centering
		\includegraphics[width=0.5\linewidth]{figures/Alphabet_G2.png}
		\caption{Alphabet G2}
		\label{fig:G2_Reg_a}
	\end{subfigure}
\par
\vspace{0.8em}
\par
\begin{subfigure}{\linewidth}
		\centering
		\includegraphics[width=0.8\linewidth]{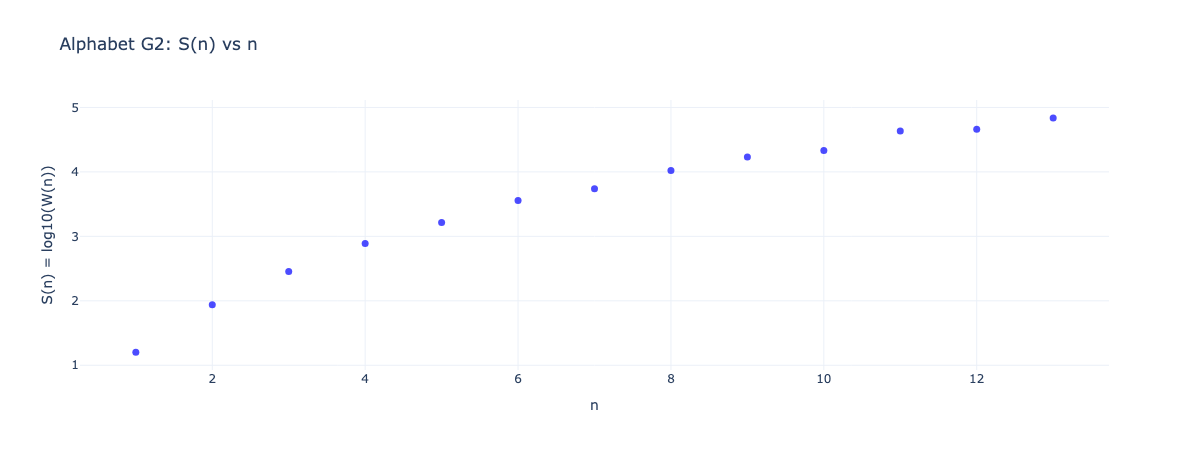}
		\caption{$S_{\Gamma}(n)$ vs $n$}
		\label{fig:G2_Reg_b}
	\end{subfigure}
\par
\vspace{0.8em}
\par
\begin{subfigure}{\linewidth}
		\centering
		\includegraphics[width=0.8\linewidth]{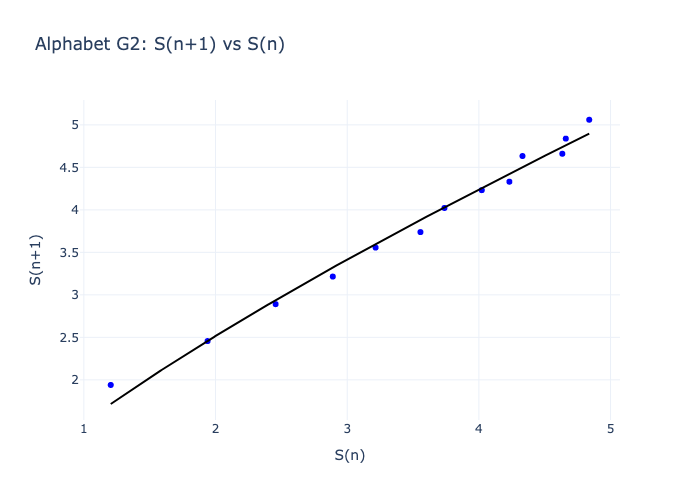}
		\caption{$S_{\Gamma}(n+1)$ vs $S_{\Gamma}(n)$}
		\label{fig:G2_Reg_c}
	\end{subfigure}
\caption{Alphabet G2 and its associated entropy plots. While the behavior is
not linear, it exhibits a monotonically increasing trend.}%
\label{fig:G2_Reg}%
\end{figure}

\begin{figure}[th]
\centering
\par
\begin{subfigure}{0.7\linewidth}
		\centering
		\includegraphics[width=0.5\linewidth]{figures/Alphabet_G3.png}
		\caption{Alphabet G3}
		\label{fig:G3_Reg_a}
	\end{subfigure}
\par
\vspace{0.8em}
\par
\begin{subfigure}{\linewidth}
		\centering
		\includegraphics[width=0.8\linewidth]{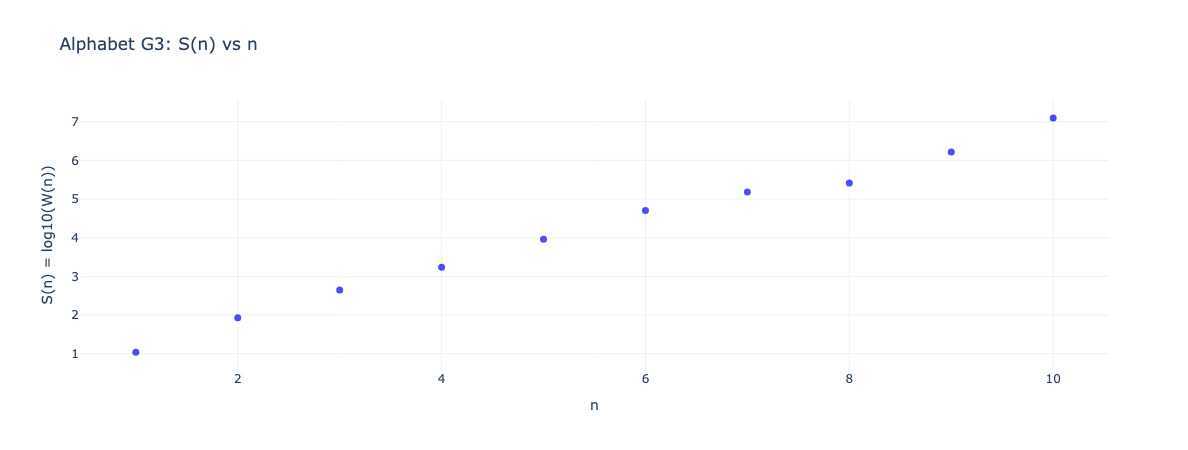}
		\caption{$S_{\Gamma}(n)$ vs $n$}
		\label{fig:G3_Reg_b}
	\end{subfigure}
\par
\vspace{0.8em}
\par
\begin{subfigure}{\linewidth}
		\centering
		\includegraphics[width=0.8\linewidth]{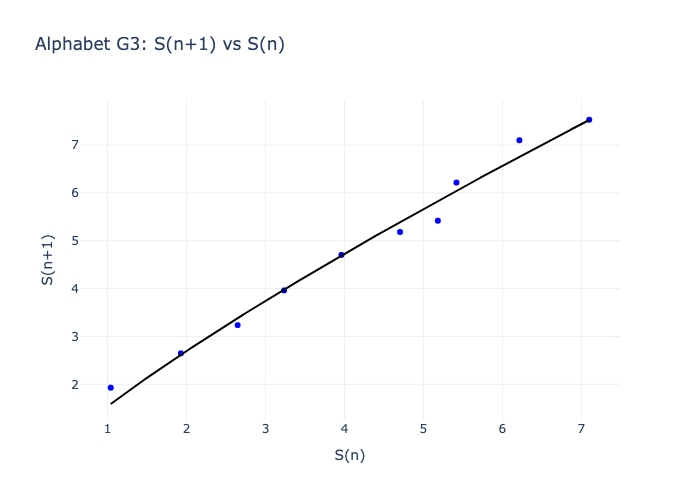}
		\caption{$S_{\Gamma}(n+1)$ vs $S_{\Gamma}(n)$}
		\label{fig:G3_Reg_c}
	\end{subfigure}
\caption{Alphabet G3 and its associated entropy plots.}%
\label{fig:G3_Reg}%
\end{figure}

\begin{figure}[th]
\centering
\par
\begin{subfigure}{0.6\linewidth}
		\centering
		\includegraphics[width=0.7\linewidth]{figures/Alphabet_B1.png}
		\caption{Alphabet B1}
		\label{fig:B1_Reg_a}
	\end{subfigure}
\par
\vspace{0.8em}
\par
\begin{subfigure}{\linewidth}
		\centering
		\includegraphics[width=0.8\linewidth]{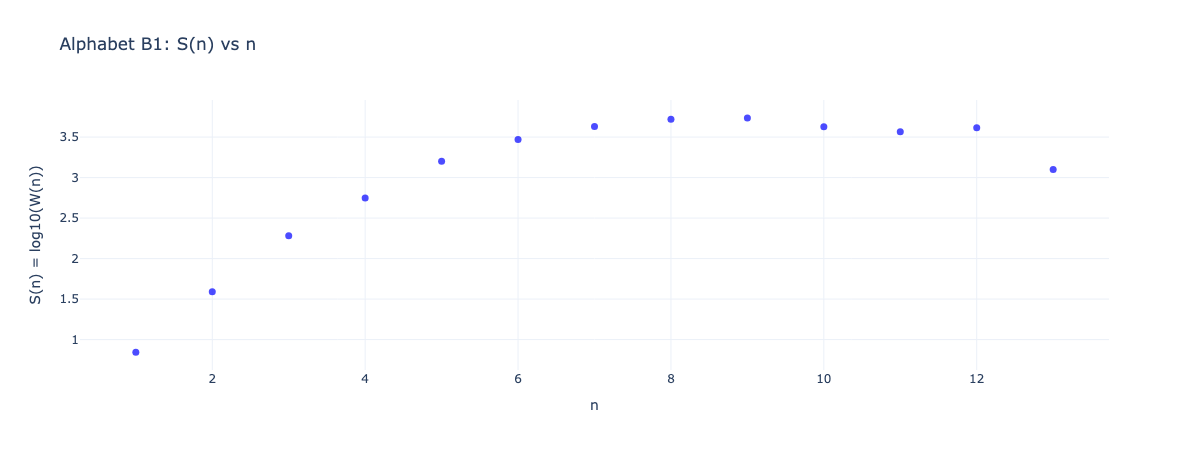}
		\caption{$S_{\Gamma}(n)$ vs $n$}
		\label{fig:B1_Reg_b}
	\end{subfigure}
\par
\vspace{0.8em}
\par
\begin{subfigure}{\linewidth}
		\centering
		\includegraphics[width=0.8\linewidth]{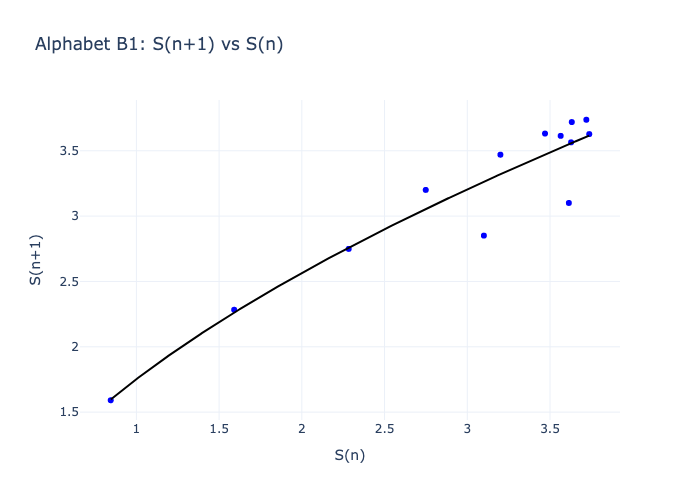}
		\caption{$S_{\Gamma}(n+1)$ vs $S_{\Gamma}(n)$}
		\label{fig:B1_Reg_c}
	\end{subfigure}
\caption{Alphabet B1 and its associated entropy plots.}%
\label{fig:B1_Reg}%
\end{figure}

\begin{figure}[th]
\centering
\par
\begin{subfigure}{0.6\linewidth}
		\centering
		\includegraphics[width=0.7\linewidth]{figures/Alphabet_B2.png}
		\caption{Alphabet B2}
		\label{fig:B2_Reg_a}
	\end{subfigure}
\par
\vspace{0.8em}
\par
\begin{subfigure}{\linewidth}
		\centering
		\includegraphics[width=0.5\linewidth]{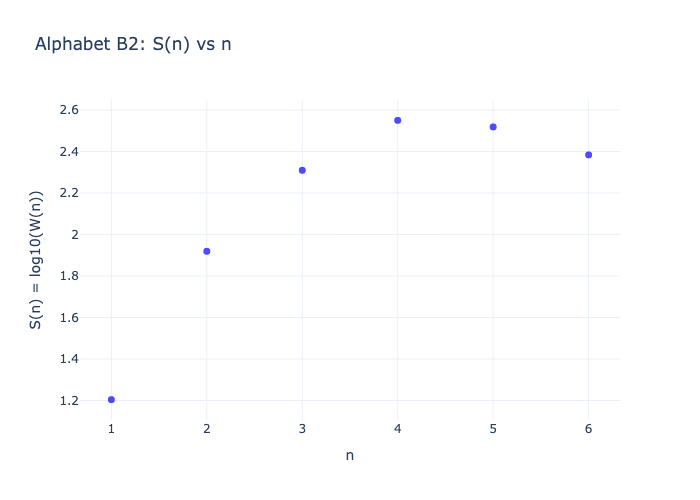}
		\caption{$S_{\Gamma}(n)$ vs $n$}
		\label{fig:B2_Reg_b}
	\end{subfigure}
\par
\vspace{0.8em}
\par
\begin{subfigure}{\linewidth}
		\centering
		\includegraphics[width=0.5\linewidth]{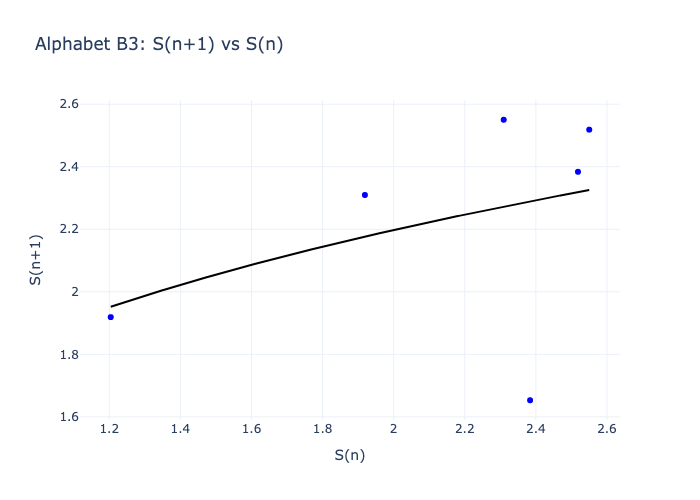}
		\caption{$S_{\Gamma}(n+1)$ vs $S_{\Gamma}(n)$}
		\label{fig:B2_Reg_c}
	\end{subfigure}
\caption{Alphabet B2 and its associated entropy plots.}%
\label{fig:B2_Reg}%
\end{figure}

\begin{figure}[th]
\centering
\par
\begin{subfigure}{0.6\linewidth}
		\centering
		\includegraphics[width=0.7\linewidth]{figures/Alphabet_B3.png}
		\caption{Alphabet B3}
		\label{fig:B3_Reg_a}
	\end{subfigure}
\par
\vspace{0.8em}
\par
\begin{subfigure}{\linewidth}
		\centering
		\includegraphics[width=0.8\linewidth]{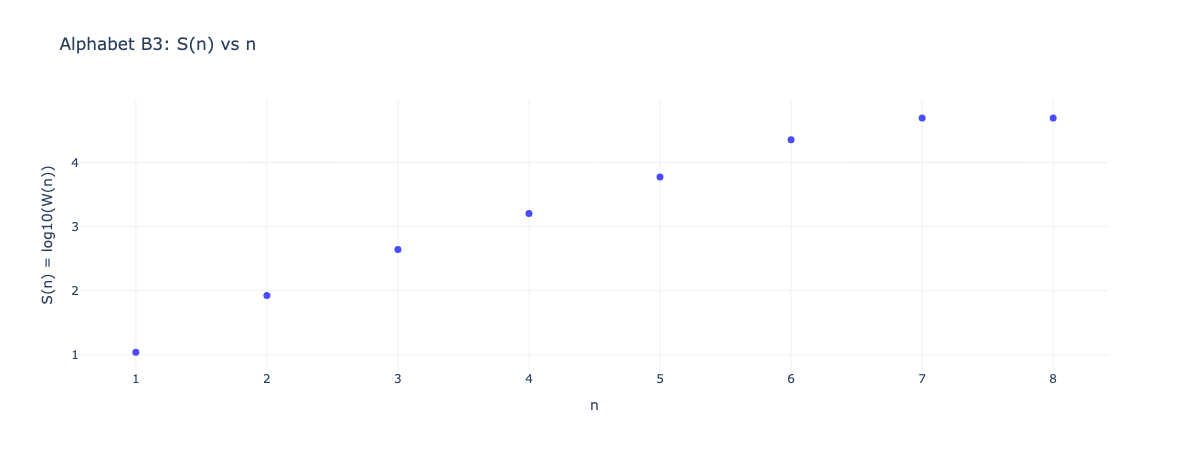}
		\caption{$S_{\Gamma}(n)$ vs $n$}
		\label{fig:B3_Reg_b}
	\end{subfigure}
\par
\vspace{0.8em}
\par
\begin{subfigure}{\linewidth}
		\centering
		\includegraphics[width=0.8\linewidth]{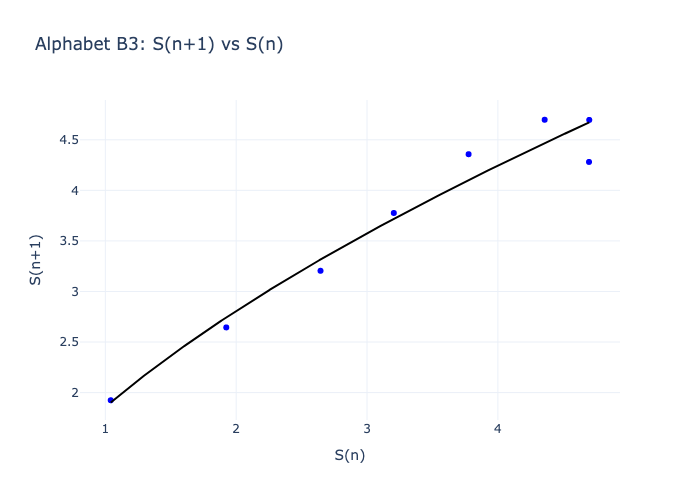}
		\caption{$S_{\Gamma}(n+1)$ vs $S_{\Gamma}(n)$}
		\label{fig:B3_Reg_c}
	\end{subfigure}
\caption{Alphabet B3 and its associated entropy plots.}%
\label{fig:B3_Reg}%
\end{figure}

To rigorously analyze the growth monotonicity of $S_{\Gamma}(n+1)$ versus
$S_{\Gamma}(n)$ as defined in Eq.(\ref{map3.1}), we perform a regression
analysis for each alphabet. Instead of Pearson's correlation coefficient, we
employ Kendall's Tau ($\tau$) to assess the relationship between data pairs.
This choice is specifically motivated by our primary focus on monotonicity
rather than strict linearity, and by the superior robustness of Kendall's Tau
to outliers. Subsequently, we compare the alphabets based on their respective
regression coefficients $c_{0}$ and $\gamma$, as well as the Kendall's Tau
coefficient ($\tau$) , as summarized in \textbf{FIG.} (\ref{fig:G1_Reg} to
\ref{fig:B3_Reg}) and \textbf{Table} (\ref{regression_results}).

\textit{For the good alphabets} (namely, the alphabets which can cover the
whole plane) $f_{\Gamma}\left(  S_{\Gamma}\left(  n\right)  \right)
=f_{\Gamma}\left(  z\right)  $ is very close to a power low (with very small
oscillations):
\begin{equation}
f_{\Gamma}\left(  z\right)  =c_{0}z^{\gamma}\left(  1+\omega(z)\right)
\ ,\ \ \gamma>0\ , \label{good}%
\end{equation}
where $c_{0}$ is a constant and $\omega(z)$ is an oscillating function with a
small amplitude:
\begin{equation}
\left\vert \omega(z)\right\vert \ll1\ . \label{good1}%
\end{equation}

\begin{table}[ptbh]
\centering
\par%
\begin{tabular}
[c]{cccc}\hline
Alphabet & $c_{0} $ & $\gamma$ & $\tau$\\\hline
G1 & 1.25 & 0.899 & 1.00\\
G2 & 1.49 & 0.754 & 1.00\\
G3 & 1.54 & 0.810 & 1.00\\
B1 & 1.75 & 0.550 & 0.72\\
B2 & 1.87 & 0.233 & 0.33\\
B3 & 1.86 & 0.596 & 0.79\\\hline
\end{tabular}
\caption{Alphabets Regression coeff}%
\label{regression_results}%
\end{table}

\textbf{FIG.}  \ref{fig:G1_Reg} presents the complexity analysis for Alphabet G1.
This alphabet demonstrates a remarkably predictable growth pattern,
characteristic of "good" alphabets, evident in both its entropy plots and
regression coefficients. \textbf{FIG.}  \ref{fig:G1_Reg}(b) shows a consistent and
monotonic increase in $S_{\Gamma}(n)$ with $n$, indicating a stable,
non-erratic growth in the number of distinct tiles. The core analysis of
growth monotonicity, as depicted in \textbf{FIG.} \ref{fig:G1_Reg}(c) plotting
$S_{\Gamma}(n+1)$ against $S_{\Gamma}(n)$, reveals an almost perfectly linear
relationship between successive entropy values. This strong linearity is
quantitatively supported by the regression coefficients from  \textbf{Table} 
\ref{regression_results}: a scaling factor $c_{0} = 1.25$ indicates a
consistent base growth rate, and an exponent $\gamma= 0.899$ (approaching
unity) signifies a nearly linear relationship in the logarithmic domain,
consistent with stable complexity evolution. Crucially, a Kendall's Tau
coefficient of $\tau= 1.00$ confirms a perfect positive monotonic
relationship, meaning $S_{\Gamma}(n+1)$ invariably increases with $S_{\Gamma
}(n)$, validating the consistent and predictable growth inherent to "good"
alphabets. This empirical evidence for minimal oscillation strongly supports
the theoretical condition $\left\vert \omega(z)\right\vert \ll1$ in Equation
(\ref{good}).

\textbf{FIG.}  \ref{fig:G2_Reg} displays another example of a "good" alphabet (G2). It
is illustrated the growth of $S_{\Gamma}(n)$ with $n$, showing a consistent,
albeit slightly less uniform, monotonic increase in entropy compared to G1.
This indicates a generally stable increase in tiles complexity. The
relationship between $S_{\Gamma}(n+1)$ and $S_{\Gamma}(n)$, depicted in \textbf{FIG.} 
\ref{fig:G2_Reg}(c), reveals a positive linear correlation, with most data
points closely adhering to the regression line. From  \textbf{Table} 
\ref{regression_results}, Alphabet G2 exhibits regression coefficients $c_{0}
= 1.49$ and $\gamma= 0.754$. While $\gamma$ is slightly lower than G1, it
still signifies a robust, near-linear growth in the logarithmic domain,
consistent with predictable complexity. More critically, the Kendall's Tau
coefficient $\tau= 1.00$ confirms a perfect positive monotonic relationship,
reinforcing that $S_{\Gamma}(n+1)$ consistently increases with $S_{\Gamma}%
(n)$. This perfect monotonicity, combined with the high linearity, categorizes
Alphabet G2 as having highly stable and desirable growth characteristics,
fully supporting the minimal oscillation condition $\left\vert \omega
(z)\right\vert \ll1$ in Equation (\ref{good}).

\textbf{FIG.} \ref{fig:G3_Reg} presents the complexity analysis for Alphabet G3,
another example of a "good" alphabet. \textbf{FIG.} \ref{fig:G3_Reg}(b) illustrates
the growth of $S_{\Gamma}(n)$ with $n$, showing a consistent monotonic
increase in entropy, indicative of stable combinatorial expansion. The core
analysis of growth monotonicity, depicted in \textbf{FIG.}  \ref{fig:G3_Reg}(c)
plotting $S_{\Gamma}(n+1)$ against $S_{\Gamma}(n)$, reveals a positive linear
correlation, with data points closely adhering to the regression line. As per
 \textbf{Table} \ref{regression_results}, Alphabet G3 exhibits regression coefficients
$c_{0} = 1.54$ and $\gamma= 0.810$. The $\gamma$ value, close to unity,
consistently signifies a robust, near-linear growth in the logarithmic domain,
characteristic of predictable complexity. Critically, the Kendall's Tau
coefficient $\tau= 1.00$ confirms a perfect positive monotonic relationship,
where $S_{\Gamma}(n+1)$ consistently increases with $S_{\Gamma}(n)$. This
perfect monotonicity, combined with the observed linearity, firmly categorizes
Alphabet G3 as having highly stable and desirable growth characteristics,
fully supporting the minimal oscillation condition $\left\vert \omega
(z)\right\vert \ll1$ in Equation (\ref{good}).

\textbf{FIG.}  \ref{fig:B1_Reg} depicts the complexity analysis for Alphabet B1,
classifying it as a "bad" alphabet. \textbf{FIG.}  \ref{fig:B1_Reg}(b) illustrates
that $S_{\Gamma}(n)$ shows an increasing trend with $n$, but with noticeable
plateaus and less uniform steps, hinting at unstable combinatorial growth. The
key divergence from "good" alphabets is evident in the $S_{\Gamma}(n+1)$ vs
$S_{\Gamma}(n)$ plot (\textbf{FIG.} \ref{fig:B1_Reg}(c)). While a general linear
trend can be observed, there is significant dispersion of data points around
the regression line, indicating structural inconsistencies and a less
predictable relationship between successive entropy values. From  \textbf{Table} 
\ref{regression_results}, Alphabet B1 has coefficients $c_{0} = 1.75$ and
$\gamma= 0.550$. The $\gamma$ value, being significantly lower than unity,
indicates a weaker, sub-linear growth in the logarithmic domain. Crucially,
the Kendall's Tau coefficient $\tau= 0.72$ is markedly less than 1.00. This
value confirms a positive correlation but explicitly indicates the presence of
monotonic inversions, reflecting the limitations in its combinatorial expansion.

The analysis for Alphabet B2, shown in \textbf{FIG.}  \ref{fig:B2_Reg}, reveals a
behavior that is distinctly more erratic than B1. Its $S_{\Gamma}(n)$ vs $n$
plot (\textbf{FIG.}  \ref{fig:B2_Reg}(b)) lacks a clear, stable increasing trend and
is characterized by significant, unpredictable scatter. This instability is
confirmed in the $S_{\Gamma}(n+1)$ vs $S_{\Gamma}(n)$ plot (\textbf{FIG.} 
\ref{fig:B2_Reg}(c)), which shows extreme dispersion and no discernible linear
pattern, indicating severe local inconsistencies. The regression coefficients
from  \textbf{Table}  \ref{regression_results} quantify this "bad" behavior. With $c_{0}
= 1.87$ and a growth coefficient $\gamma= 0.233$, the growth is substantially
weaker than B1's and approaches zero. More significantly, the Kendall's Tau
coefficient $\tau= 0.33$ is exceptionally low. This value indicates an almost
complete breakdown of monotonic correlation, highlighting profound structural
disorder and a stark contrast to the more structured, albeit flawed, growth
seen in B1.

\textbf{FIG.}  \ref{fig:B3_Reg} presents the complexity analysis for Alphabet B3,
which also falls into the category of "bad" alphabets. \textbf{FIG.} \ref{fig:B3_Reg}%
(b) illustrates the behavior of $S_{\Gamma}(n)$ with $n$. Similar to B1 and
B2, while an overall increasing trend is observed, the growth is not perfectly
smooth, showing some irregularities in the rate of entropy increase. More
critically, the $S_{\Gamma}(n+1)$ vs $S_{\Gamma}(n)$ plot in \textbf{FIG.} 
\ref{fig:B3_Reg}(c) demonstrates significant deviation from a strict linear
relationship, exhibiting a noticeable scatter of data points around the
regression line. This scatter is particularly pronounced at higher $S_{\Gamma
}(n)$ values, indicating a less predictable and less stable progression of
complexity. From  \textbf{Table}  \ref{regression_results}, Alphabet B3 has regression
coefficients $c_{0} = 1.86$ and $\gamma= 0.596$. The $\gamma$ value,
significantly less than unity, confirms a sub-linear growth in the logarithmic
domain, implying that the incremental increase in complexity diminishes with
increasing sequence length, a hallmark of "bad" alphabets. The most telling
characteristic is the Kendall's Tau coefficient $\tau= 0.79$, which, while
higher than B1/B2, remains significantly below 1.00. This value definitively
indicates the presence of monotonic violations, where $S_{\Gamma}(n+1)$ does
not consistently increase with $S_{\Gamma}(n)$, reflecting the inherent
structural inconsistencies and the non-negligible oscillations of $\omega(z)$
in Equation (\ref{good}).

The comprehensive analysis of alphabets, based on their entropy growth
$S_{\Gamma}(n)$, reveals a clear distinction in their combinatorial behavior.
The results for "bad" alphabets, exemplified by B1, B2, and B3, consistently
indicate a sub-linear growth in entropy (characterized by $\gamma< 1$) and a
loss of perfect monotonicity (indicated by $\tau< 1.00$). This behavior is
consistent with an inherent limitation in their ability to efficiently tile
the entire space, often stemming from the absence of superposable structures.
This diminished growth rate and structural inconsistency is precisely
corroborated by the specific $c_{0}$, $\gamma$, and $\tau$ values presented in
 \textbf{Table}  \ref{regression_results}, and visibly reinforced by the corresponding
$S_{\Gamma}(n+1)$ vs $S_{\Gamma}(n)$ plots. Conversely, "good" alphabets (G1,
G2, G3) exhibit near-linear growth in the logarithmic domain ($\gamma\approx
1$) and perfect monotonicity ($\tau= 1.00$), demonstrating stable and
predictable complexity expansion. Consequently, these empirical observations
provide a robust basis for distinguishing between alphabets with \textit{good}
and \textit{bad} tiling properties based on their entropy scaling and
monotonic characteristics.

\textit{For bad alphabets}, the numerical fits of the plots can still be
described by Eq. (\ref{good}). On the other hand, the condition in Eq.
(\ref{good1}) is violated for bad alphabets, namely the irregular oscillations
around the power law behavior are large.

Thus, the \textit{partial decidability protocols} proposed in the present
manuscript is based on this pronounced difference. In simple terms, the
protocol reads as follows:

\textbf{Step 1)} Compute numerically as many values as possible of the
function $W_{\Gamma}\left(  n\right)  $ (subject to the available resources,
the ideal situation being having $n_{\max}$ larger than $q$ of, at least, a
factor of 2 or 3: see the comments below Eq. (\ref{protocol1a})) for the
alphabet $\Gamma$ of interest.

\textbf{Step 2)} Due to the large amount of combinations, it is better to use
the entropy $S_{\Gamma}\left(  n\right)  $ as main variable .

\textbf{Step 3)} Produce the plot $P=(X,Y)=\left(  S_{\Gamma}\left(  n\right)
,S_{\Gamma}\left(  n+1\right)  \right)  $ as described in the previous section.

\textbf{Step 4)} Construct the best fit of $S_{\Gamma}\left(  n+1\right)
=f_{\Gamma}\left(  S_{\Gamma}\left(  n\right)  \right)  $ and compute the
employ Kendall's Tau ($\tau$).

\section{Physical interpretation in terms of discrete Chaos}

As it has been already emphasized in the introduction, there is a deep
connections between deterministic chaos and undecidability since the Chaitin's
results in \cite{15a0}. The "coexistence" of determinism and undecidability
(see, for instance, the detailed analysis in \cite{15a} \cite{15d} \cite{15e})
implies that undecidability can manifest itself in dynamical systems
manifesting chaotic behavior. The idea to use the second protocol instead of
the first (simpler) protocol is that the second one reveals interesting
informations on the transition from good to bad alphabets. The qualitative
explanation of the effectiveness of the present approach is based on the
transition to chaos in discrete mappings of logistic type (see, for a detailed
review on discrete chaos, \cite{[15]}).

Let us interpret
\begin{equation}
S_{\Gamma}\left(  n+1\right)  =f_{\Gamma}\left(  S_{\Gamma}\left(  n\right)
\right)  \ , \label{dynsyst1}%
\end{equation}
as a discrete dynamical system (a possible choice discussed in the previous
section is$\ f_{\Gamma}\left(  z\right)  =c_{0}z^{\gamma}\left(
1+\omega(z)\right)  $ although the results in the present sections apply to
generic form of $f_{\Gamma}\left(  z\right)  $).

In particular, if $f_{\Gamma}\left(  z\right)  $ \textit{is monotone
increasing\footnote{According to the present numerical results, this
corresponds (when the parametrization $f_{\Gamma}\left(  z\right)
=c_{0}z^{\gamma}\left(  1+\omega(z)\right)  $ is used) to the case in which
$\left\vert \omega(z)\right\vert \ll1$.} (namely, }$\frac{d}{dz}f_{\Gamma
}\left(  z\right)  >0$\textit{)} the solutions of the above dynamical system
will tend to the solutions of the simpler (non-chaotic) dynamical system here
below:
\begin{equation}
S_{\Gamma}\left(  n+1\right)  =f_{\Gamma}^{(0)}\left(  S_{\Gamma}\left(
n\right)  \right)  \ ,\ \ \ f_{\Gamma}^{(0)}\left(  z\right)  =c_{0}z^{\gamma
}\ , \label{dynsyst2}%
\end{equation}
according to which $W_{\Gamma}\left(  n\right)  $ grows exponentially with $n$
with subexponential corrections.

On the other hand, if $f_{\Gamma}\left(  z\right)  $ \textit{is not monotone
increasing\footnote{According to the present numerical results, this
corresponds (when the parametrization $f_{\Gamma}\left(  z\right)
=c_{0}z^{\gamma}\left(  1+\omega(z)\right)  $ is used) to the case in which
$\left\vert \omega(z)\right\vert $ \textit{is not small compared to 1}.}
(namely, }$\frac{d}{dz}f_{\Gamma}\left(  z\right)  $ vanishes and changes
sign, generically, more than once\textit{) and local maxima and minima
appear}, the dynamical system in Eq. (\ref{dynsyst1}) can enter into a chaotic
phase. In particular, the more peaked is the local maximum, the closer is the
shape of $f_{\Gamma}\left(  z\right)  $ to a logistic map in the chaotic
phase. Thus, although the explicit analytic form of $f_{\Gamma}\left(
z\right)  $ is not available, the present results show that the maps
$f_{\Gamma}\left(  z\right)  $ associated to good alphabets do not manifest
chaotic tendency while the ones associated to bad alphabets manifest a clear
chaotic tendency.

In particular, in the terminal region of the experimental data for non-tiling
("bad") alphabets, one can observe a departure from sustained growth,
manifesting as a decreasing trend in the relationship between successive
values. While initial iterations may exhibit increasing values of $S(n+1)$
with respect to $S(n)$, the final experimental points reveal a clear
inversion, indicating that larger values of $S(n)$ are associated with smaller
subsequent values of $S(n+1)$. This decay suggests an inherent constraint
within these alphabets that limits their expansive behavior.

This eventual decrease in $S(n+1)$ as $S(n)$ increases can be qualitatively
described by a functional form reminiscent of a downward-opening parabola.
Although not necessarily an exact fit, a quadratic model of the type
\begin{equation}
S(n+1)\approx S_{0}-a[S(n)-b]^{2}\ ,
\end{equation}
where $S_{0}$, $a>0$, and $b$ are constants, captures the essence of this
behavior. This form illustrates how, beyond a certain point related to $b$,
increasing $S(n)$ leads to a reduction in $S(n+1)$, ultimately resulting in
the observed decay. This tendency towards a decreasing relationship at larger
values of $S(n)$ appears to be a distinguishing feature of alphabets that are
unable to tile the plane\footnote{In this respect, it would be very
interesting to construct a family $\Gamma$ \textit{which does not satisfy the
present heuristics and which, nevertheless, can cover the full plane}.}: the
similarity with the logistic map is manifest. We will come back on this very
interesting issue in a future publication.

\section{Quantifying "goodness" and "badness" of alphabets}

In this section, we will discuss how the physical arguments in the previous
sections are actually confirmed by a statistical analysis of the numerical
data. In complex systems involving sequential or time-dependent data,
quantifying the relative quality of different datasets presents significant
analytical challenges. This section examines how correlation measures can
serve as effective tools for evaluating monotonic relationships within our
alphabets, allowing us to distinguish between what we term \textquotedblleft
good" and \textquotedblleft bad" behaviors.

When analyzing correlation in datasets with extreme values and seeking to
identify monotonic trends rather than strictly linear relationships, the
choice of correlation coefficient becomes crucial. While the Pearson
correlation coefficient is commonly employed in statistical analysis, its
limitations make it suboptimal for our specific requirements.

The Pearson correlation coefficient ($r$) measures the linear relationship
between variables:%

\[
r = \frac{\text{cov}(X, Y)}{\sigma_{X} \sigma_{Y}}%
\]

However, this measure presents several drawbacks for our analysis. Pearson's
coefficient assumes linearity in relationships, is highly sensitive to extreme
values and outliers, requires normally distributed data, and can be
significantly influenced by the scale and magnitude of observations. These
characteristics make it inadequate for our datasets, which contain extreme
values and where we are primarily concerned with monotonic trends rather than
linear relationships.

Kendall's Tau ($\tau$), a non-parametric measure of rank correlation, offers a
more suitable alternative for our analysis:%

\[
\tau= \frac{(\text{Number of concordant pairs}) - (\text{Number of discordant
pairs})}{\frac{1}{2} n (n - 1)}%
\]

Kendall's Tau evaluates correlation based on the concordance of
pairs---whether the ranks of paired observations move in the same
direction---rather than the magnitude of differences. This approach provides
several advantages for our specific analytical needs:

First, Kendall's Tau is particularly robust when dealing with datasets
containing outliers or extreme values, as is the case with several of our
alphabets. Because it focuses on the relative ordering of data points rather
than their actual values, the coefficient is less affected by unusually large
or small observations.

Second, Kendall's Tau excels at detecting monotonic relationships without
assuming linearity. Our primary interest lies in determining whether the
regression patterns across different alphabets consistently increase or
decrease, making monotonic behavior more relevant than strict linearity.

Third, the coefficient maintains its reliability even with smaller sample
sizes, which is beneficial when analyzing rarer alphabetic patterns.

Kendall's Tau has demonstrated its utility across numerous scientific
disciplines. In environmental science, researchers have employed it to detect
monotonic trends in climate data where extreme weather events might skew other
correlation measures \cite{[16]}. In economics, it has proven valuable for
analyzing market behaviors when dealing with volatile price movements
\cite{[17]}. Researchers regularly use Kendall's Tau when examining
relationships between particulate matter progression, particularly when
distributions are non-normal \cite{[18]}.

Our methodology involves applying Kendall's Tau to measure the monotonic
trends present in different alphabets. For ``good" alphabets, we observe high
$\tau$ values, indicating strong monotonic relationships consistent with
theoretical expectations. Conversely, ``bad" alphabets exhibit lower $\tau$
values, reflecting weaker monotonic behavior and greater dispersion in the data.

By employing Kendall's Tau as our primary correlation measure, we establish a
robust framework for distinguishing between alphabets that exhibit consistent,
predictable behavior (high $\tau$ values) and those that demonstrate erratic
patterns (low $\tau$ values). This approach provides an objective metric for
quantifying the relative \textquotedblleft goodness" or \textquotedblleft
badness" of different alphabets, facilitating more rigorous comparative
analysis and more reliable conclusions about their underlying properties.

\section{Resuming the results of the present work}

Here we will resume what we have done in the present work. We have associated
to any alphabet $\Gamma$ a temperature, an entropy and a partition function.
The requirement of a good thermodynamical behavior provides one with a very
effective heuristic able to identify good candidates $\Gamma$ to tile the
plane. The simplest of all is the first heuristic which only uses the fact
that in most of physically reasonable systems the degeneracy of the energy
level is an increasing function of the energy. The second heuristic which has
been defined discloses a nice analogy between the transition from good to bad
alphabets and the transition from regular to chaotic behavior in discrete
mapping of logistic type. Despite the fact that neither of those heuristics
are sufficient conditions for tilability, they are very effective and disclose
quite intriguing relations between statistical mechanics, chaos theory and
decidability in mathematics and physics. In order to use these ideas in
practice, the Kendall's Tau must be used. It seems that the present strategy
has a quite wide range of possible applications.

\subsection{What we have not done}

First of all, it is important to remark that we have proposed a very effective
heuristic that is able to identify whether or not a given alphabet is a good
candidate to tile the whole $%
%TCIMACRO{\U{211d} }%
%BeginExpansion
\mathbb{R}
%EndExpansion
^{2}$. The word "heuristic" means that there will always be families that are
classified as "good" but that do not tile the plane. The way to achieve these
"\textit{fake good alphabets}" is to add some additional bits on the tiles to
pump-up the entropy in such a way to make examples where the number of tilings
of $n\times n$ squares grows exponentially until a given $n_{\max}$ but the
tiles still fail to tile the plane\footnote{We thank the referee for this
remark.}. Another way to state this problem is the following. In practice one
cannot compute $W_{\Gamma}\left(  n\right)  $ for any $n$ so that one has to
stop at some finite $n_{\max}$. We do not have a rigorous answer to the
question of how large $n_{\max}$ should be (compared to $q$) in order to be
confident that the alphabet under examination is a good candidate to tile the
plane. For instance, we cannot exclude the following situation. We have the
numerical verification of Eq. (\ref{map1.9}) for ($1$, $2$, $3$,..,
$q\cdot10^{10^{q}}$) but then, when $n=1+q\cdot10^{10^{q}}$, the condition in
Eq. (\ref{map1.9}) is violated (thus, in such a case, we would be looking at a
fake good alphabet whose "fakiness" manifests itself at a value of $n$ which
is too large to be detected by any existing machine). It is worth emphasizing
here that if a given family $\Gamma$ fails to cover up the whole $%
%TCIMACRO{\U{211d} }%
%BeginExpansion
\mathbb{R}
%EndExpansion
^{2}$ then, necessarily, at some point Eq. (\ref{map1.9}) must be violated. On
the other hand, our results strongly suggest that these "\textit{fake good
alphabets}" do not easily show up in experiments since with our heuristic we
have been able to identify successfully all the available examples of good
tilings in the literature. Moreover, it seems that if $n_{\max}$ is 2 or 3
times $q$, then $n_{\max}$ is large enough to be confident that the alphabet
is a good candidate to tile the plane.

As far as the similarity of the transition from good to bad alphabets and of
the transition from regular to chaotic behavior in discrete mapping of
logistic types, we have not shown rigorously that bad alphabets are always
associated to chaotic discrete mappings while good alphabet to regular ones.
The difficulty lies in the fact that the analytic form of $f_{\Gamma}\left(
z\right)  $ cannot be computed and so one has to use numerical fits of the
mapping $f_{\Gamma}\left(  z\right)  $. The presence or absence of chaotic
behavior may depend on the family of functions which are chosen to do the fit.
However, the connections between deterministic chaos and undecidability
(discussed in details in \cite{15a0} \cite{15a} \cite{15d} \cite{15e} and
references therein) lead to the fact that the existence of deterministic chaos
in dynamical systems can be related to a sort of Godelian undecidability
rather than to the typical numerical intractability. Since the Wang tiling
problem is also undecidable, it is reasonable to assume that discrete mappings
associated to Wang alphabets manifest their undecidability through chaotic
behavior (consistently with \cite{15a} \cite{15d} \cite{15e}). It is also
worth emphasizing that the relation with chaotic mapping could also be used to
obtain a "negative heuristic". Namely, if the discrete mapping of our
construction associated to a given family $\Gamma$ is chaotic until $n_{\max}$
(with $n_{\max}$ large enough compared to $q$), then it is unlikely that one
could prove that such $\Gamma$ tiles the whole plane, even if it does.

We will come back on these interesting issues in a future publication.

\section{Conclusions and perspectives}

In the present manuscript we have proposed two partial decidability protocols
(two \textit{heuristics}) for the Wang tilings problem (which is the prototype
of the undecidable problem in combinatorics and statistical mechanics). The
idea is to define effective entropy and temperature associated to any alphabet
$\Gamma$ (together with the corresponding partition function). A subclass of
\textit{good alphabets} can be identified by requiring, basically, a good
thermodynamical behavior: such good alphabets are good candidates to tile the
whole plane (if such heuristics are satisfied until a large enough $n_{\max}%
$). This proposal has been tested successfully with the known available good
alphabets (which produce periodic tilings, aperiodic but self-similar tilings
as well as tilings which are neither periodic nor self-similar). From the
theoretical physics viewpoint, it is a very intriguing result to be able to
produce effective heuristics using sound arguments from statistical mechanics.
The present analysis also shows that the transition from good to bad alphabets
is very similar to the transition from regular to chaotic behavior in discrete
mappings of logistic type (and this fact could also be used to define negative
heuristics, as it has been emphasize in the previous section).

From the practical viewpoint, the present results are quite powerful. The hard
computational challenges encountered in scaling Wang tilings, particularly the
rapid increase in effort required to expand the tiling plane with larger
alphabets, are related to the fact that the Wang tilings problem is
\textit{co-RE complete} (the first level in the arithmetical hierarchy). The
observed distinction between "good" alphabets (which are good candidates to
tile the whole plane) and "bad" alphabets (which exhibit irregular
oscillations and eventual decreasing trends in their associated mappings)
further hints at a fundamental difference in their complexity. Of course, this
study does not resolve the P vs NP problem. Nevertheless, the apparent
difficulty in efficiently finding a tiling solution, even when a valid
configuration can be readily verified, aligns with the prevailing conjecture
that P $\neq$ NP, implying that the Wang tiling problem could represent a
scenario where finding a solution is significantly more computationally
demanding than checking its correctness (at least in the case of bad alphabets).

\subsection*{Acknowledgements}

The authors would like to thank the anonymous referee for valuable suggestions
and discussions. This work has been funded by Fondecyt grants No. 1240048,
1240043, 1240247 and by Grant ANID EXPLORACIÓN 13250014\textbf{. }The Centro
de Estudios Cientificos (CECs) is funded by the Chilean Government through the
Centers of Excellence Base Financing Program of Conicyt.


\begin{thebibliography}{99}                                                                                               %


\bibitem {[1]}H. Wang, Proving theorems by pattern recognition. II, Bell Syst.
Tech. J. 40 (1961) 1--42.

\bibitem {[2]}R. Berger, The Undecidability of the Domino Problem, Mem. Amer.
Math. Soc., vol.66, 1966.

\bibitem {[3]}R. Robinson, Undecidability and nonperiodicity of tilings in the
plane, Invent. Math. 12 (1971) 177--209.

\bibitem {[4]}E. Jeandel, M. Rao, An Aperiodic Set of 11 Wang Tiles, Advances
in Combinatorics, 2021:1, 37 pp.

\bibitem {[4.1]}S. Labbé., A Self-Similar Aperiodic Set of 19 Wang Tiles,
Geometriae Dedicata 201 (2019), pp. 81--109.

\bibitem {[4.2]}J. Kari, A Small Aperiodic Set of Wang Tiles, Discrete
Mathematics 160 (1996), pp. 259--264.

\bibitem {[5]}D. Shechtman, I. Blech, D. Gratias, J. Cahn, Metallic Phase With
Long-Range Orientational Symmetry and No Translational
Symmetry\textquotedblright, Phys. Rev. Lett. 53, 1951--1953 (1984).

\bibitem {[6]}L. Bindi, N. Yao, C. Lin, L. Hollister, C. Andronicos, V.
Distler, M. Eddy, A. Kostin, V. Kryachko, G. MacPherson, W. Steinhardt, M.
Yudovskaya, P. Steinhardt, Natural Quasicrystal With Decagonal Symmetry,
Scientific Reports 5, 9111 (2015).

\bibitem {[7]}D. Levine, P. Steinhardt, Quasicrystals: A New Class of Ordered
Structures, Phys. Rev. Lett. 53 (26 1984), pp. 2477--2480.

\bibitem {[8.1]}J. Mikekisz, A Microscopic Model With Quasicrystalline
Properties, Journal of Statistical Physics 58.5--6 (1990), pp. 1137--1149;
Stable Quasicrystalline Ground States, Journal of Statistical Physics 88.3--4
(1997), pp. 691--711; An Ultimate Frustration in Classical Lattice-Gas Models,
Journal of Statistical Physics 90.1--2 (1998), pp. 285--300.

\bibitem {[9]}A. van Enter, W. Ruszel, Chaotic Temperature Dependence at Zero
Temperature, Journal of Statistical Physics 127.3 (2007), pp. 567--573. 10.1007/s10955-006-9260-2.

\bibitem {[10]}Jean-René Chazottes, M. Hochman, On the Zero-Temperature Limit
of Gibbs States, Communications in Mathematical Physics 297.1 (2010), pp.
265--281. 10.1007/s00220-010-0997-8.

\bibitem {[11]}Leo Gayral. Complexity and Robustness of Tilings with Random
Perturbations, PhD thesis (Université Paul Sabatier - Toulouse III, 2023) https://theses.hal.science/tel-04288597/.

\bibitem {[11.1]}Á. Perales-Eceiza, T. Cubitt, M. Gu, D. Pérez-García, M. M.
Wolf, Undecidability in Physics: a Review, https://doi.org/10.48550/arXiv.2410.16532.

\bibitem {[13]}T. S. Cubitt, D. Perez-Garcia, M. M. Wolf, M.M., Undecidability
of the spectral gap. Nature 528, 207 (2015); Comment on \textquotedblleft on
the uncomputability of the spectral gap\textquotedblright. arXiv:1603.00825.

\bibitem {[14]}N. Shiraishi, K. Matsumoto, Undecidability in quantum
thermalization, Nature Communications 12, 5084 (2021).

\bibitem {[15]}S. N. Elaydi, Discrete Chaos: With Applications in Science and
Engineering, Chapman, Hall (CRC Press 2007).

\bibitem {15a0}G. J. Chaitin, \textit{Scientific American} \textbf{232} (5),
47 (1975); \textit{Int. I. Theor, Phys} \textbf{22}, 941 (1982).

\bibitem {15a}M. O. Rabin, \textit{Trans. Amer. Math. Soc.} \textbf{141}
(1969), 1-35.

\bibitem {15d}C. Agnes, M. Rasetti, \textit{Nuovo Cimento} \textbf{106}, 879
(1991); \textit{Chaos, Solitons \& Fractals} \textbf{5}, 161-175 (1995).

\bibitem {15e}M. Rasetti, \textit{Chaos, Solitons \& Fractals} \textbf{5},
133-138 (1995).

\bibitem {[16]}Burn, D. H. and Elnur, M. A. H. (2002), Detection of hydrologic
trends and variability. Journal of Hydrology, 255(1-4), 107-122.

\bibitem {[17]}Croux, C., and Dehon, C. (2010), Influence functions of the
Spearman and Kendall correlation measures. Statistical Methods, Applications,
19(4), 497-515.

\bibitem {[18]}Hernandez, W., Mendez, A., Zalakeviciute, R., and Diaz-Marquez,
A., M., (2020). Analysis of the Information Obtained from PM2.5 Concentration
Measurements in an Urban Park, https://doi.org/10.1109/TIM.2020.2966360

\bibitem {[19]}Grünbaum, B., and Shephard, G. C. (1987). Tilings and patterns.
W. H. Freeman and Company.

\bibitem {[20]}Pak, I., and Yang, J. (2013). Tiling simply connected regions
with rectangles. Journal of Combinatorial Theory, Series A, 120(7), 1804-1816. https://doi.org/10.1016/j.jcta.2013.06.008.

\bibitem {[21]}Aamand, A., Abrahamsen, M., Ahle, T. D., and Rasmussen, P. M.
R. (2023). Tiling with squares and packing dominos in polynomial time. ACM
Transactions on Algorithms, 19(3), Article 30, 1--28. https://doi.org/10.1145/3597932

\bibitem {[22]}Arora, S., and Barak, B. (2009). Computational complexity: A
modern approach. Cambridge University Press.
\end{thebibliography}
\end{document}